\documentclass[10.5pt,compsoc]{tst}
\usepackage{graphicx}
\usepackage{footmisc}
\usepackage{subfigure}
\usepackage{url}
\usepackage{multirow}
\usepackage[noadjust]{cite}
\usepackage{amsmath,amsthm}
\usepackage{amssymb,amsfonts}
\usepackage{algorithm}
\usepackage{algorithmic}
\usepackage{booktabs}
\usepackage{bm}
\usepackage{color}
\usepackage{ccaption}
\usepackage{booktabs}
\usepackage{float}
\usepackage{fancyhdr}
\usepackage{caption}
\usepackage{xcolor,stfloats}
\usepackage{comment}
\graphicspath{{figures/}}
\usepackage{cuted}  
\usepackage{captionhack}
\usepackage{epstopdf}
\usepackage{setspace}

\headevenname{\zihao{-5}{\textbf{\emph{Tsinghua Science and Technology, February}} 2018, 23(1): 000--000}}%
\headoddname{{\sf Zhenglin Huang et al.:}\quad {\textbf{\emph{A Simple But Accurate Approximation for Multivariate Gaussian Rate-Distortion Function ...}}}}%

\newtheoremstyle{mystyle}{0pt}{0pt}{\normalfont}{1em}{\bf}{}{1em}{}
\theoremstyle{mystyle}

\newtheorem{lemma}{\textbf{Lemma}}
\newtheorem{theorem}{\textbf{Theorem}}

\newtheorem{corollary}{\textbf{Corollary}}
\newcommand{\nop}[1]{}

\newcommand{\tr}{{\rm tr}}
\newcommand{\diag}{{\rm diag}}
\newcommand{\mean}{{\rm mean}}
\definecolor{red}{rgb}{1,0,0}
\definecolor{green}{rgb}{0,1,0}
\definecolor{blue}{rgb}{0,0,0}

\theoremstyle{remark}
\newtheorem{remark}{Remark}

\makeatletter
\renewcommand{\@biblabel}[1]{[#1]\hfill}
\makeatother

\begin{document}

\hyphenpenalty=50000
\allowdisplaybreaks

\makeatletter
\newcommand\mysmall{\@setfontsize\mysmall{7}{9.5}}

\newenvironment{tablehere}
  {\def\@captype{table}}
  {}
\newenvironment{figurehere}
  {\def\@captype{figure}}
  {}

\thispagestyle{plain}%
\thispagestyle{empty}%

\let\temp\footnote
\renewcommand \footnote[1]{\temp{\zihao{-5}#1}}
{}
\vspace*{-40pt}
\noindent{\normalsize\textbf{\scalebox{0.885}[1.0]{\makebox[5.9cm][s]
{TSINGHUA\, SCIENCE\, AND\, TECHNOLOGY}}}}

\vskip .2mm
{\normalsize
\textbf{
\hspace{-5mm}
\scalebox{1}[1.0]{\makebox[5.6cm][s]{%
I\hfill S\hfill S\hfill N\hfill{\color{white}%
l\hfill l\hfill}1\hfill0\hfill0\hfill7\hfill-\hfill0\hfill2\hfill1\hfill4
\hfill \color{white}{\quad 0\hfill ?\hfill /\hfill ?\hfill ?\quad p\hfill p\hfill  ?\hfill ?\hfill ?\hfill --\hfill ?\hfill ?\hfill ?}\hfill}}}}

\vskip .2mm
{\zihao{5-}
\textbf{
\hspace{-5mm}
\scalebox{1}[1.0]{\makebox[5.6cm][s]{%
DOI:~\hfill~\hfill1\hfill0\hfill.\hfill2\hfill6\hfill5\hfill9\hfill9\hfill/\hfill T\hfill S\hfill T\hfill.\hfill2\hfill0\hfill1\hfill8\hfill.\hfill9\hfill0\hfill1\hfill0\hfill0\hfill0\hfill0}}}}


\begin{strip}
{\center
{\zihao{3}\textbf{
A Simple but Accurate Approximation for Multivariate Gaussian Rate-Distortion Function and Its Application in Maximal Coding Rate Reduction
\vskip 1.2mm\noindent
}}
\vskip 9mm}

{\center {\sf \zihao{5}
Zhenglin Huang, Qifa Yan$^*$, Bin Dai, and Xiaohu Tang\\
}
\vskip 5mm}
%

\centering{
\begin{tabular}{p{160mm}}

{\zihao{-5}
\linespread{1.6667} %
\noindent
\bf{Abstract:} {\sf
The multivariate Gaussian rate-distortion (RD) function is crucial in various applications, such as digital communications, data storage, or neural networks. However, the complex form of the multivariate Gaussian RD function prevents its application in many neural network-based scenarios that rely on its analytical properties, for example, white-box neural networks, multi-device task-oriented communication, and semantic communication. This paper proposes a simple but accurate approximation for the multivariate Gaussian RD function. The upper and lower bounds on the approximation error (the difference between the approximate and the exact value) are derived, which indicate that for well-conditioned covariance matrices, the approximation error is small. In particular, when the condition number of the covariance matrix approaches 1, the approximation error approaches 0. In addition, based on the proposed approximation, a new classification algorithm called Adaptive Regularized ReduNet (AR-ReduNet) is derived by applying the approximation to ReduNet, which is a white-box classification network oriented from Maximal Coding Rate Reduction (MCR$^2$) principle. Simulation results indicate that AR-ReduNet achieves higher accuracy and more efficient optimization than ReduNet.}
\vskip 4mm
\noindent
{\bf Key words:} {\sf Maximal Coding Rate Reduction, Multivariate Gaussian Distribution, Neural Networks, Rate-Distortion}}

\end{tabular}
}
\vskip 6mm

\vskip -3mm
\zihao{6}\end{strip}

\thispagestyle{plain}%
\thispagestyle{empty}%
\makeatother
\pagestyle{tstheadings}

\begin{figure}[b]
\vskip -6mm
\begin{tabular}{p{44mm}}
\toprule\\
\end{tabular}
\vskip -2.5mm
\noindent
\setlength{\tabcolsep}{1pt}
\begin{tabular}{p{1.5mm}p{79.5mm}}
$\bullet$& Zhenglin Huang, Qifa Yan, Bin Dai, and Xiaohu Tang are with School of Information Science and Technology, Southwest Jiaotong University, Chengdu, 610031, China, and Information Coding and Transmission Key Laboratory of Sichuan Province, CSNMT Int. Coop. Res. Centre (MoST), Southwest Jiaotong University, Chengdu, 611756, China. E-mail: zlhuang@my.swjtu.edu.cn; qifayan@swjtu.edu.cn; daibin@swjtu.edu.cn; xhutang@swjtu.edu.cn. (\emph{Corresponding Author: Qifa Yan})\\
$\bullet $&This paper was supported in part by the National Key R\&D Program of
China under Grant 2022YFA1005000, in part by the National Natural Science
Foundation of China under Grants 62101464 and 12141108. \\
$\sf{*}$
          &          Manuscript received: 2024-08-19;
          revised: 2024-10-11;
          accepted: 2024-11-14
\end{tabular}
\end{figure}\zihao{5}

\vspace{3.5mm}
\zihao{5}
\section{Introduction}
\textcolor{blue}{Rate-distortion (RD) theory is used to characterize the trade-off between rate and distortion in lossy compression, based on a series of Shannon's seminal works \cite{shannon1948mathematical, shannon1959coding}}. \textcolor{blue}{Over the past several decades, RD theory has been actively studied in the field of information theory \cite{berger2003rate}.} In some typical \textcolor{blue}{applications}, the RD function is widely employed in \textcolor{blue}{source coding \cite{ortega1998rate, sullivan1998rate}, video stream rate allocation \cite{ji2004optimal}, and compressed sensing \cite{coluccia2014operational, leinonen2018rate}}. Recently, RD theory has been applied to more fields, especially machine learning, e.g., lossy compression based on deep learning \cite{agustsson2019generative, choi2019variable, hu2020improving, li2021learning}, steganography \cite{li2024high}, \textcolor{blue}{Maximal Coding Rate Reduction (MCR$^2$)} principle \cite{yu2020learning, chan2022redunet, tong2022incremental, dai2022ctrl}, rate-distortion-perception trade-off \cite{blau2019rethinking}, and semantic information theory \cite{liu2022indirect, stavrou2023role, lyu2024semantic}. To better evaluate the performance of the algorithms in these applications, it is necessary to obtain the analytical expressions of RD functions.

Fortunately, when the distributions of sources are certain typical distributions, such as Gaussian distribution, Bernoulli distribution, etc., the analytical expression of the RD function is known \cite{cover1999elements}. In particular, the Gaussian distribution is the most important in practice. For example, Gaussian distribution is usually used to build mathematical models \cite{gopinath1998maximum, zhang2018cellular, dang2022efficient}. In machine learning, the Variational Auto Encoder (VAE) assumes that hidden variables are Gaussian random variables \cite{kingma2019introduction}. In addition, in the MCR$^2$ principle, the RD function of Gaussian vectors is used as the objective function to derive ReduNet \cite{chan2022redunet}.

\begin{figure}[tbp]
\centering
\includegraphics[scale=0.55]{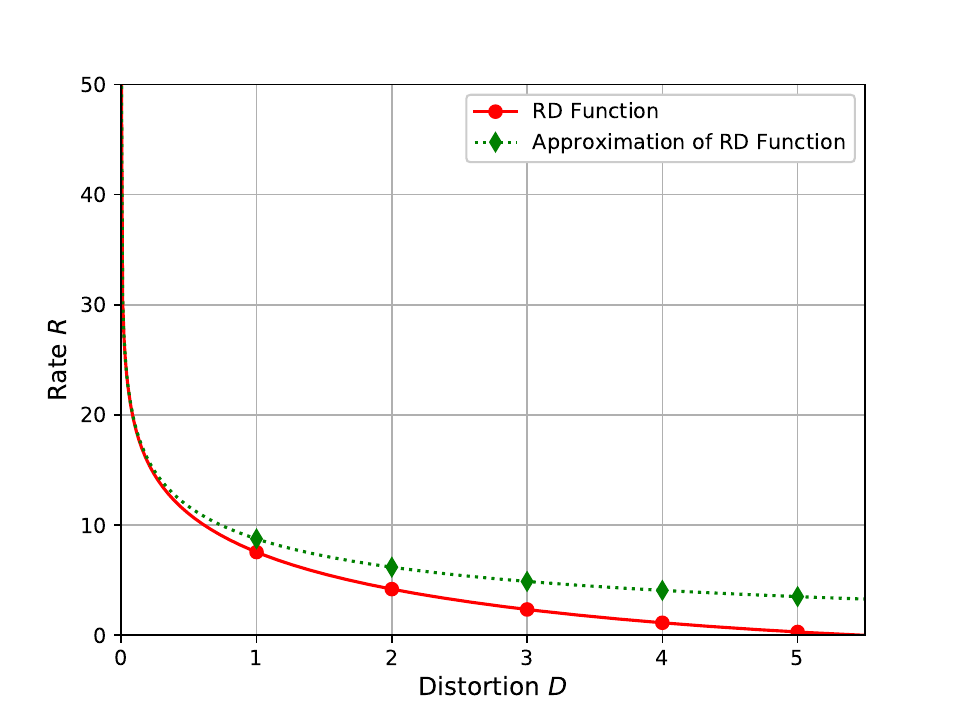}
\caption{The exact multivariate Gaussian RD function and its approximation in Ref. \cite{ma2007segmentation}, for the covariance matrix $\mathbf{\Sigma}=\text{diag}(1.0, 0.9, 0.8, 0.7, 0.6, 0.5, 0.4, 0.3, 0.2, 0.1) $.}
\label{fig:introduction_fig}
\end{figure}

Although the expression of the multivariate Gaussian RD function is known, it does not have good analytical properties, e.g., the form of the derivative is complex. In many applications (e.g., ReduNet), it is desirable to use these analytical properties. In order to alleviate this difficulty, a simple explicit function was proposed and utilized as an approximation of the multivariate Gaussian RD function in Ref. \cite{ma2007segmentation}. However, as illustrated in Fig. \ref{fig:introduction_fig}, the gap between the approximation in Ref. \cite{ma2007segmentation} and the exact RD function increases with the distortion $ D $. This means that the approximation of the RD function in Ref. \cite{ma2007segmentation} is only accurate when the value of $ D $ is close to $ 0 $.

This motivates us to find a more accurate approximation of the exact multivariate Gaussian RD function. The main contributions of this paper are:
\begin{enumerate}
\item To approximate the $n$-dimensional multivariate Gaussian RD function $R(D)$, a simple function is chosen from a family of convex functions $\{R_{\alpha}(D):\alpha\in[0,1]\}$, where the parameter $\alpha$ is chosen such that $ R_{\alpha}(D) $ attains $0$ at the same value with $ R(D) $. It turns out that there exists a unique value $\alpha^*$ satisfying the constraint, which can be easily obtained by a bisection algorithm. In particular, whenever the condition number of the covariance matrix approaches $1$, the per-dimensional difference $\frac{R_{\alpha^*}(D)-R(D)}{n}$ approaches $0$ for all values of $D$ in the interested interval.

\item Based on the above approximation, Adaptive Regularized ReduNet (AR-ReduNet) is proposed by incorporating the approximation into the framework of ReduNet, which is a white-box classification network oriented to the MCR$^2$ principle. Simulation results on the MNIST dataset indicate that AR-ReduNet increases the classification accuracy, due to the improved approximation for the RD function.
\end{enumerate}

The remainder of this paper is organized as follows: Section \ref{sec:preliminaries} reviews the necessary preliminaries for the computation of the multivariate Gaussian RD function. Section \ref{sec:results} presents the parametric approximation function and theoretical bounds, Section \ref{sec:bounds} discusses the choice of the parameter and numerical results, Section \ref{sec:applications} derives the AR-ReduNet and presents simulation results, and Section \ref{sec:conclusions} concludes this paper.

\section{Preliminaries} \label{sec:preliminaries}
Consider \textcolor{blue}{a random variable $\bm X=(X_1,\ldots,X_n)^{\text{T}}\in\mathbb{R}^n$} from some continuous distribution \textcolor{blue}{$p(\bm x)=\text{\rm Pr}(\bm X=\bm x)$}. \textcolor{blue}{Here, $ \bm x=(x_1,\ldots,x_n)^{\text{T}}\in\mathbb{R}^n $ denotes an observed value of the random variable $ \bm X $.} The RD function with mean square distortion measure for a given distortion $D\in\mathbb{R}^+$ is defined by
\begin{align}
\textcolor{blue}{R(D)=\mathop{\inf}_{p(\bm{\hat x}|\bm{x}):\mathbb{E}[||\bm{X}-\bm{\hat X}||^2]\leq D}I\big(\bm{X};\bm{\hat X}\big),}
\end{align}
where \textcolor{blue}{${\bm{\hat X}}\in\mathbb{R}^n$} is the estimation of \textcolor{blue}{$\bm X$} generated from the conditional probability distribution $p(\bm{\hat x}|\bm x)$, \textcolor{blue}{where $ \bm{\hat x} $ denotes an observed value of $ \bm{\hat X} $.} $\mathbb{E}[\cdot]$ is the expectation over the joint distribution $p({\bm x}, {\bm {\hat x}})=p(\bm x)p(\bm{\hat x}|\bm  x)$, and $I(\cdot;\cdot)$ is the mutual information. Essentially, the function $R(D)$ characterizes the optimal average coding length of the random vector $\bm x$ under the constraint that the average mean square error \textcolor{blue}{${\mathbb{E}}[||\bm X-\bm{\hat X} ||^2]$} is less than $D$ \cite{cover1999elements}.

In practical machine learning applications, the most widely used source is the Gaussian source \cite{chan2022redunet, tong2022incremental, dai2022ctrl, yu2023white}. For Gaussian vectors\footnote{The mean can be any vector in $\mathbb{R}^n$.} with the covariance matrix \textcolor{blue}{$ \bm \Sigma$}, the exact RD function is given by (see Ref. \cite{cover1999elements} for details):
    \begin{align}
    R(D)=\sum_{i=1}^{n}\frac{1}{2}\log\frac{\lambda_i}{D_{i}},\label{eq:RD}
    \end{align}
where $\lambda_1,\ldots,\lambda_n$ are eigenvalues of the matrix \textcolor{blue}{$\bm \Sigma$},
\begin{subequations} \label{eq:Di}
  \begin{align}
    D_{i}=\left\{\begin{array}{cc}
    L, &{\rm{if}}~L < \lambda_i, \\
    \lambda_i, &{ \rm{if}}~L \geq \lambda_i
    \end{array}\right.,
    \end{align}
and $ L $ is chosen to satisfy
\begin{align}
\sum_{i=1}^{n}D_{i}=D.
\end{align}
\end{subequations}

In particular, when $0< \frac{D}{n}\leq \lambda_{\min}=\min \{\lambda_{1},\ldots, \lambda_{n}\}$, it turns out that $ D_i=L=\frac{D}{n} $, and $ R(D) $ is given by
\begin{align}
R_0(D)\triangleq\frac{1}{2}\log \det\Big(\frac{n}{D}\textcolor{blue}{\bm \Sigma}\Big).\label{eq:smallD}
\end{align}
That is, for $ 0 < D\leq n\lambda_{\min} $, the exact RD function can be computed directly from \eqref{eq:smallD}. Nevertheless, for $ D > n\lambda_{\min} $, the expression in \eqref{eq:smallD} is no longer accurate. It was proposed in Ref. \cite{ma2007segmentation} to use
\begin{align}
    R_{1}(D)\triangleq\frac{1}{2}\log \det\left(\textcolor{blue}{\bm I}+\frac{n}{D}\textcolor{blue}{\bm \Sigma}\right)  \label{RID}
\end{align}
to approximate $R(D)$ over $D\in(0,+\infty)$. Recently, the function $R_1(D)$ was used in ReduNet to replace the exact RD function \cite{chan2022redunet}. However, as illustrated in Fig. \ref{fig:introduction_fig}, the function $R_1(D)$ is only accurate when $ D $ is small. \textcolor{blue}{Let $\tr(\bm \Sigma)$ be the trace of $\bm \Sigma$, i.e, the sum of all diagonal elements.} By \eqref{eq:Di}, $R(D)$ attains $0$ when $D=\sum_{i=1}^n\lambda_i=\tr(\textcolor{blue}{\bm \Sigma})$, while the value of $R_1(D)$ at $D={\rm tr}(\textcolor{blue}{\bm \Sigma})$ is
\begin{align}\label{eqn:R1}
R_1({\tr}(\textcolor{blue}{\bm \Sigma}))=\frac{1}{2} \log\det\Big(\textcolor{blue}{\bm I}+\frac{n}{{\tr}(\textcolor{blue}{\bm \Sigma})}\textcolor{blue}{\bm \Sigma}\Big)>0.
\end{align}
From Fig. \ref{fig:introduction_fig}, it can be observed that $R_1(D)$ shows significant discrepancies with $ R(D) $, when $ D\rightarrow\tr(\textcolor{blue}{\bm \Sigma}) $.

\section{A Parameterized Approximation Family} \label{sec:results}
As the exact function $R(D) $ is equal to $0$ when $ D \geq \tr(\textcolor{blue}{\bm \Sigma}) $, the interested interval to approximate $ R(D) $ is $[0, \tr(\textcolor{blue}{\bm \Sigma})]$. The idea to improve the approximation over $ R_1(D) $ is to add a multiplicative scalar $\alpha\in[0,1]$ to the regularization term $\textcolor{blue}{\bm I}$ in \eqref{RID}. That is, define
\begin{align}
R_\alpha(D)=\frac{1}{2}\log\det \Big(\alpha \textcolor{blue}{\bm I}+\frac{n}{D}\textcolor{blue}{\bm \Sigma}\Big), \ \forall\, \alpha\in[0,1].\label{eqn:RalpD}
\end{align}

For each $\alpha\in[0,1]$, $R_\alpha(D)$ is a convex function defined on $(0,+\infty)$. Notice that $ R_\alpha(D) $ degrades to $R_0(D)$ and $R_1(D)$ when $\alpha=0$ and $1$ respectively. The following theorem gives the bounds on the difference $\frac{R_{\alpha}(D)-R(D)}{n}$ for each $ \alpha \in [0, 1] $:

\begin{theorem}\label{thm:1}
For any $\alpha\in[0,1]$, the approximation $ R_{\alpha}(D) $ and the exact RD function $ R(D) $ satisfy:
\begin{subequations}\label{bounds}
\begin{itemize}
\item If $0<D\leq n\lambda_{\min}$,
\begin{align}
0\leq \frac{R_{\alpha}(D)-R(D)}{n}\leq \frac{1}{2}\log(1+\alpha). \label{bound:1}
\end{align}
\item If $n\lambda_{\min}< D\leq \tr(\textcolor{blue}{\bm \Sigma})$,
\begin{align}
\frac{1}{2}\log(1+\alpha) - \frac{1}{2}\log &\frac{\lambda_{\mean}}{\lambda_{\min}}  \leq \frac{R_{\alpha}(D) - R(D)}{n} \nonumber \\
& \leq \frac{1}{2}\log(1+\alpha). \label{bound:2}
\end{align}
where $ \lambda_{\mean}=(\lambda_1+ \ldots +\lambda_n)/n=\tr(\textcolor{blue}{\bm \Sigma})/n.  $
\end{itemize}
\end{subequations}
\end{theorem}

\begin{IEEEproof}
Since \textcolor{blue}{$ \bm \Sigma$} is a non-negative symmetric matrix, it has the following form:
\begin{align}
\textcolor{blue}{\bm \Sigma=\bm U\bm \Lambda \bm U^{\rm T},}\label{eq:decom}
\end{align}
where \textcolor{blue}{$\bm U$} is an orthogonal matrix and \textcolor{blue}{$\bm \Lambda=\diag(\lambda_1,\ldots,\lambda_n)$} is the diagonal matrix with its eigenvalues $\lambda_1,\ldots,\lambda_n$ at the diagonal positions. Therefore,
\begin{align}
R_\alpha(D)&=\frac{1}{2}\log\det\Big(\alpha \textcolor{blue}{\bm I}+\frac{n}{D}\textcolor{blue}{\bm U \bm \Lambda \bm U^{\rm T}}\Big)\\
&=\frac{1}{2}\log \det \Big(\textcolor{blue}{\bm U}\Big(\alpha \textcolor{blue}{\bm I}+\frac{n}{D}\textcolor{blue}{\bm \Lambda}\Big)\textcolor{blue}{\bm U^{\rm T}}\Big)\\
&=\frac{1}{2}\log \det(\textcolor{blue}{\bm U})\det\Big(\alpha \textcolor{blue}{\bm I}+\frac{n}{D}\textcolor{blue}{\bm \Lambda}\Big)\det(\textcolor{blue}{\bm U^{\rm T}}) \\
&=\frac{1}{2}\log \det\Big(\alpha \textcolor{blue}{\bm I}+\frac{n}{D}\textcolor{blue}{\bm \Lambda}\Big)\\
&=\frac{1}{2}\log \prod_{i=1}^n \Big(\alpha+\frac{n\lambda_i}{D}\Big).\label{eq:RaDexpression}
\end{align}
Notice that by \eqref{eq:Di},
$ D_i=\min \{ L, \lambda_i \} $. Together with \eqref{eq:RaDexpression} and \eqref{eq:RD},
\begin{subequations}
\begin{align}
&R_{\alpha}(D)-R(D) \nonumber \\
=&\sum_{i=1}^{n}\frac{1}{2}\log\Big(\alpha+\frac{n\lambda_{i}}{\sum_{j=1}^{n}\min\{L, \lambda_{j}\}}\Big)\notag\\
&\quad\quad\quad\quad-\sum_{i=1}^{n}\frac{1}{2}\log\Big(\frac{\lambda_{i}}{\min\{L, \lambda_{i}\}}\Big)\notag\\
=&\sum_{i=1}^{n}\frac{1}{2}\log\Big[\Big(\alpha+\frac{n\lambda_{i}}{\sum_{j=1}^{n}\min\{L, \lambda_{j}\}}\Big)  \nonumber \\
&\quad\quad\quad\quad\Big(\frac{\min\{L, \lambda_{i}\}}{\lambda_{i}}\Big)\Big]    \nonumber \\
=&\sum_{i=1}^{n}\frac{1}{2}\log\Big(\frac{\alpha \min\{L, \lambda_{i}\}}{\lambda_i} \nonumber \\
&\quad\quad\quad\quad+\frac{\min\{L, \lambda_{i}\}}{\frac{1}{n}\sum_{j=1}^{n}\min\{L, \lambda_{j}\}}\Big)   \label{eq:step} \\
=&\frac{n}{2}\sum_{i=1}^{n}\frac{1}{n}\log\Big(\frac{\alpha \min\{L, \lambda_i\}}{\lambda_i} \nonumber \\
&\quad\quad\quad\quad+\frac{\min\{L, \lambda_i\}}{\frac{1}{n}\sum_{j=1}^{n}\min\{L, \lambda_j\}}\Big)   \notag \\
\leq&\frac{n}{2}\log\Big(\frac{\alpha}{n}\sum_{i=1}^{n}\frac{\min\{L, \lambda_i\}}{\lambda_i} \nonumber \\
&\quad\quad\quad\quad+\frac{\frac{1}{n}\sum_{i=1}^{n}\min\{L, \lambda_i\}}{\frac{1}{n}\sum_{j=1}^{n}\min\{L, \lambda_j\}}\Big) \label{explain:a} \\
    =&\frac{n}{2}\log\Big(\frac{\alpha}{n}\sum_{i=1}^{n}\frac{\min\{L, \lambda_i\}}{\lambda_i}+1\Big)  \nonumber \displaybreak[0] \\
    \leq&\frac{n}{2}\log(1+\alpha),   \label{UPPERBOUND}
\end{align}
\end{subequations}
where $\eqref{explain:a}$ follows from the fact that the function $ \log\left(\alpha x+\frac{y}{\frac{1}{n}\sum_{j=1}^{n}\min\{L, \lambda_j\}}\right) $ is concave in the variables $(x,y)$. Thus, the upper bound in \eqref{bounds} is proved.

For the lower bound, consider the case $0<D\leq n\lambda_{\min}$, by \eqref{eq:smallD},
\begin{align}
R(D)=\sum_{i=1}^n\frac{1}{2}\log\Big(\frac{n\lambda_i}{D}\Big).
\end{align}
Thus, by \eqref{eq:RaDexpression},
\begin{align}
R_{\alpha}(D)-R(D)=\sum_{i=1}^n\frac{1}{2}\log\Big(1+\frac{\alpha D}{n\lambda_i}\Big)\geq 0.\label{lower:1}
\end{align}

For the case $n\lambda_{\min}< D\leq \tr(\textcolor{blue}{\bm \Sigma})$, by \eqref{eq:step},
\begin{subequations}
\begin{align}
&R_{\alpha}(D)-R(D) \nonumber \\
=&\sum_{i=1}^{n}\frac{1}{2}\log\Big(\frac{\alpha }{\lambda_i}+\frac{1}{\frac{1}{n}\sum_{j=1}^{n}\min\{L, \lambda_j\}}\Big)\notag\\
&\quad\quad\quad+\sum_{i=1}^n\frac{1}{2}\log \min\{L, \lambda_i\}\\
=&\frac{n}{2}\sum_{i=1}^n\frac{1}{n}\log \Big(\frac{\alpha }{\lambda_i}+\frac{1}{\frac{1}{n}\sum_{j=1}^{n}\min\{L, \lambda_j\}}\Big)\notag\\
&\quad\quad\quad+\sum_{i=1}^n\frac{1}{2}\log \min\{L, \lambda_i\}\\
\geq&\frac{n}{2}\log\Big(\frac{\alpha}{\frac{1}{n}\sum_{i=1}^n\lambda_i}+\frac{1}{\frac{1}{n}\sum_{j=1}^n\min\{L,\lambda_j\}}\Big)\notag\\
&\quad\quad\quad+\sum_{i=1}^n\frac{1}{2}\log \min\{L, \lambda_i\}\label{exp:111}\\
\geq& \frac{n}{2}\log\Big(\frac{\alpha}{\frac{1}{n}\sum_{i=1}^n\lambda_i}+\frac{1}{\frac{1}{n}\sum_{j=1}^n \lambda_j}\Big)\notag\\
&\quad\quad\quad+\sum_{i=1}^n\frac{1}{2}\log \lambda_{\min} \\
=&\frac{n}{2}\log\Big(\frac{\alpha}{\lambda_{\mean}}+\frac{1}{\lambda_{\mean}}\Big)+\frac{n}{2}\log \lambda_{\min}\\
=&\frac{n}{2}\log(1+\alpha)-\frac{n}{2}\log\frac{\lambda_{\mean}}{\lambda_{\min}}, \label{lower:2}
\end{align}
\end{subequations}
where \eqref{exp:111} follows from the convexity of the function  $\log\big(\frac{\alpha}{x}+\frac{1}{\frac{1}{n}\sum_{j=1}^n\min\{L,\lambda_j\}}\big)$ in $x$.

Finally, by \eqref{lower:1} and \eqref{lower:2}, the lower bound in \eqref{bounds} is proved.
\end{IEEEproof}

\begin{remark}\label{remark:1}
Theorem \ref{thm:1} bounds the per-dimensional difference $\frac{R_{\alpha}(D)-R(D)}{n}$ with the parameter $\alpha$ and the eigenvalues of \textcolor{blue}{$\bm \Sigma$}. In particular,
\begin{itemize}
\item Let $\alpha=0$, the upper bound in \eqref{bounds} indicates that $R_0(D)$ is actually an underestimate for $R(D)$.
\item Let $\alpha=1$, consider the case $\frac{\lambda_{\mean}}{\lambda_{\min}}<2$, the lower bound in \eqref{bound:2} is positive. Thus, the lower bound in \eqref{bounds} indicates that $R_1(D)$ is an overestimate of $ R(D) $ in this case.
\end{itemize}
\end{remark}

Theorem \ref{thm:1} and Remark \ref{remark:1} indicate that the parameter $ \alpha $ can regulate the approximation accuracy of the function $ R_{\alpha}(D) $. In the following section, we proceed to obtain an accurate approximation by selecting an appropriate value of $ \alpha $ over the interval $ [0, 1] $.

\section{Parameter Choice and Its Properties} \label{sec:bounds}
In this section, we present a method for choosing $\alpha$ and discuss the properties of the resultant approximation.

\subsection{Choice of the Parameter $\alpha$}  \label{sec:choice}
Since $ R(\tr(\textcolor{blue}{\bm \Sigma}))=0 $, the interested interval is $(0, \tr(\textcolor{blue}{\bm \Sigma)}]$. Here, the parameter $ \alpha $ is chosen to align the value of $ R_{\alpha}(D) $ with $ R(D) $ at $ D=\tr(\textcolor{blue}{\bm \Sigma}) $, that is, choose $ \alpha $ such that
\begin{align}
R_{\alpha}(\tr(\textcolor{blue}{\bm \Sigma}))=R(\tr(\textcolor{blue}{\bm \Sigma}))=0. \label{eq:condition}
\end{align}

The following lemmas provide useful guidance for searching $\alpha$.
\begin{lemma}\label{lem:property_1} For each $D>0$, the function $R_{\alpha}(D)$ increases with $\alpha$ over the interval $ [0, \infty) $.
\end{lemma}
\begin{IEEEproof}
The property is obvious by \eqref{eq:RaDexpression}.
\end{IEEEproof}

\begin{lemma}\label{lem:property_2} The function $R_{\alpha}(\tr(\textcolor{blue}{\bm \Sigma}))$ has the following properties:
\begin{enumerate}
\item $ R_{0}({\tr}(\textcolor{blue}{\bm \Sigma}))\leq 0$;
\item $ R_{1}({\tr}(\textcolor{blue}{\bm \Sigma}))> 0$.
\end{enumerate}
\end{lemma}
\begin{IEEEproof}
By \eqref{eq:decom},
\begin{align}
\tr(\textcolor{blue}{\bm \Sigma})&=\tr(\textcolor{blue}{\bm \Lambda})\\
&=\lambda_1+\lambda_2+\ldots+\lambda_n.
\end{align}
Thus, from \eqref{eq:RaDexpression},
\begin{align}
R_0(\tr(\textcolor{blue}{\bm \Sigma}))&=\frac{1}{2}\log \prod_{i=1}^n \Big(\frac{\lambda_i}{(\lambda_1+\ldots+\lambda_n)/n}\Big)\\
&=\frac{n}{2}\log\frac{\sqrt[n]{\lambda_1\lambda_2\ldots\lambda_n}}{(\lambda_1+\lambda_2+\ldots+\lambda_n)/n}\\
&\leq0,
\end{align}
where the last step follows from the fact that the geometric mean does not exceed the arithmetic mean for non-negative real numbers. This proves the property 1).

Moreover, by \eqref{eq:RaDexpression},
 \begin{align}
R_{1}(\tr(\textcolor{blue}{\bm \Sigma}))&=\frac{1}{2}\log \prod_{i=1}^n \Big(1+\frac{\lambda_i}{(\lambda_1+\ldots+\lambda_n)/n}\Big) \nonumber \\
&>0.
\end{align}
This proves the property 2).
\end{IEEEproof}

Since $ R_{\alpha}(D) $ is continuous in $\alpha$ for each $ D $, Lemma \ref{lem:property_2} indicates that there exists $ \alpha \in [0, 1]$ such that $ R_{\alpha}(\tr(\textcolor{blue}{\bm \Sigma}))=0 $. Moreover, by Lemma \ref{lem:property_1}, the value of such $\alpha$ is unique. Let $\alpha^{*}$ be the unique scalar such that $R_{\alpha^{*}}(\tr(\textcolor{blue}{\bm \Sigma}))=0$.

\textcolor{blue}{
\begin{remark}\label{remark:ex}
For any given $ \alpha \in [0, 1] $, $ R_{\alpha}(D) $ results in a continuous and differentiable approximation function. Intuitively, we choose $ \alpha=\alpha^* $ that minimizes $ |R_{\alpha}(D)-R(D)| $ at $ D=\mathrm{tr}(\bm \Sigma) $. In fact, extensive numerical evaluations indicate that the approximation error $ |R_{\alpha}(D)-R(D)| $ is uniformly small over $ D \in (0, \mathrm{tr}(\bm \Sigma)] $. As a partial support for choosing $ \alpha^* $, it can be shown that $ \alpha^* $ also minimizes the average absolute error at $n\lambda_{\min}$ and $\mathrm{tr}(\bm \Sigma)$, i.e., $ \frac{1}{2}|R_{\alpha}(n\lambda_{\min})-R(n\lambda_{\min})|+\frac{1}{2}|R_{\alpha}(\mathrm{tr}(\bm \Sigma))-R(\mathrm{tr}(\bm \Sigma))| $.
\end{remark}
}

\subsection{Properties of $R_{\alpha^*}(D)$}
The following lemma gives a bound for $ \alpha^{*} $.
\begin{lemma}\label{lem:property_alpha} The parameter $\alpha^*$ satisfies
        \begin{align}
        0\leq \alpha^{*} \leq 1-\frac{\lambda_{\min}}{\lambda_{\mean}}.  \label{alpha:bounds}
        \end{align}
\end{lemma}
\begin{IEEEproof} The lower bound $ \alpha^{*} \geq 0 $ is obvious. In the following, we prove the upper bound. By \eqref{eq:RaDexpression},
\begin{align}
R_{\alpha^{*}}(\tr(\textcolor{blue}{\bm \Sigma}))
&=\sum_{i=1}^n\frac{1}{2}\log\Big(\alpha^*+\frac{\lambda_i}{\tr(\textcolor{blue}{\bm \Sigma})/n}\Big)\\
&=\sum_{i=1}^n\frac{1}{2}\log\Big(\alpha^*+\frac{\lambda_i}{\lambda_{\mean}}\Big)\\
&\geq\sum_{i=1}^n\frac{1}{2}\log\Big(\alpha^*+\frac{\lambda_{\min}}{\lambda_{\mean}}\Big)\\
&=\frac{n}{2}\log\Big(\alpha^*+\frac{\lambda_{\min}}{\lambda_{\mean}}\Big).
\end{align}
With the fact $ R_{\alpha^{*}}(\tr(\textcolor{blue}{\bm \Sigma}))=0 $, we conclude $\alpha^*+\frac{\lambda_{\min}}{\lambda_{\mean}}\leq 1$. Thus, the upper bound in \eqref{alpha:bounds} is proved.
\end{IEEEproof}

Based on Lemma \ref{lem:property_alpha}, the following theorem gives the bounds on per-dimensional approximation error.
\begin{theorem}\label{coro:alpha}  The approximation function $R_{\alpha^{*}}(D)$ coincide with $ R(D) $ at the end points of the interval $(0, \tr(\textcolor{blue}{\bm \Sigma})]$ in the sense that\footnote{The notation $ \lim_{D\rightarrow 0^{+}}$ denotes the right limit of the function at $0$.}
\begin{align}
\lim_{D\rightarrow 0^{+}}\big(R_{\alpha^*}(D)-R(D)\big)&=0,\label{aligneqn:a}\\
R_{\alpha^*}(\tr(\textcolor{blue}{\bm \Sigma}))=R(\tr({\textcolor{blue}{\bm \Sigma}}))&=0,\label{aligneqn:b}
\end{align}
Moreover, for each $D \in (0,\tr(\textcolor{blue}{\bm \Sigma})]$,
\begin{align}
\frac{1}{2}\log\frac{\lambda_{\min}}{\lambda_{\mean}} &\leq \frac{R_{\alpha^{*}}(D)-R(D)}{n} \nonumber \\
&\leq \frac{1}{2}\log\Big(2-\frac{\lambda_{\min}}{\lambda_{\mean}}\Big). \label{BOUND_ED_alpha}
\end{align}
\end{theorem}
\begin{IEEEproof} By the choice of $\alpha^*$, \eqref{aligneqn:b} is satisfied automatically. To prove \eqref{aligneqn:a}, by \eqref{eq:Di}, when $D \leq n\lambda_{\min}$, $L=\frac{D}{n}$. This makes $L\rightarrow0^{+}$, when $D\rightarrow0^{+}$. Thus, for any $\alpha\in[0,1]$ by \eqref{eq:step},
\begin{align}
&\lim_{D\rightarrow 0^{+}} \big(R_{\alpha}(D)-R(D)\big)\\
&=\lim_{L\rightarrow0^{+}}\sum_{i=1}^{n}\frac{1}{2}\log\Big(\frac{\alpha \min\{L, \lambda_i\}}{\lambda_i} \notag\\
&\quad\quad\quad\quad\quad\quad+\frac{\min\{L, \lambda_i\}}{\frac{1}{n}\sum_{j=1}^{n}\min\{L, \lambda_j\}}\Big)\notag\\
&=\lim_{L\rightarrow 0^{+}}\sum_{i=1}^n\frac{1}{2}\log\Big(\frac{\alpha L}{\lambda_i}+1\Big)\\
&=0.
\end{align}
Therefore, the multiplicative scalar $\alpha^*$ simultaneously satisfies \eqref{aligneqn:a} and \eqref{aligneqn:b}. Then, the bounds in \eqref{BOUND_ED_alpha} directly follow from Theorem \ref{thm:1}, Lemma \ref{lem:property_alpha}, and the fact $\frac{\lambda_{\min}}{\lambda_{\mean}}\leq 1$.
\end{IEEEproof}

Let $ \kappa=\frac{\lambda_{\max}}{\lambda_{\min}} $ be the condition number of the covariance matrix, where $\lambda_{\max}=\max \{\lambda_1,\ldots,\lambda_n\} $. The following
corollary bounds the approximation error with $\kappa$.

\begin{corollary} \label{corollary:1}For each $D \in (0, \tr(\textcolor{blue}{\bm \Sigma})] $, the approximation function $R_{\alpha^*}(D)$ satisfies
\begin{align} \label{kappa}
        \frac{1}{2}\log\frac{1}{\kappa} \leq \frac{R_{\alpha^{*}}(D)-R(D)}{n} \leq \frac{1}{2}\log\left(2-\frac{1}{\kappa}\right).
        \end{align}
\end{corollary}
\begin{IEEEproof} The conclusion directly follows from Theorem \ref{coro:alpha} by the fact that $\kappa=\frac{\lambda_{\max}}{\lambda_{\min}} \geq \frac{\lambda_{\mean}}{\lambda_{\min}}$.
\end{IEEEproof}

The matrix $ \textcolor{blue}{\bm \Sigma} $ is a well-conditioned matrix when the condition number $ \kappa \rightarrow 1 $, and $ \textcolor{blue}{\bm \Sigma }$ is an ill-conditioned matrix when $ \kappa \rightarrow \infty $ \cite{horn2012matrix}. According to Corollary \ref{corollary:1}, when $\kappa \rightarrow 1 $, both the upper and lower bounds of $ \frac{R_{\alpha^{*}}(D)-R(D)}{n} $ are close to 0. That is, $ R_{\alpha^*}(D) $ is a good approximation of $R(D)$ for a well-conditioned matrix $\textcolor{blue}{\bm \Sigma}$, while for an ill-conditioned matrix, the lower bound of $ \frac{R_{\alpha^{*}}(D)-R(D)}{n} $ tends to negative infinity. In the next section, the approximation accuracy of $ R_{\alpha^{*}}(D) $ will be further investigated through numerical evaluations.

In the remainder of this section, we first describe a simple and efficient algorithm for selecting $ \alpha^{*} $, and then provide numerical results on the approximation accuracy of $R_{\alpha^*}(D)$ for well-conditioned and ill-conditioned matrices.

\subsection{A Simple and Efficient Method to Choose $ \alpha^{*} $} \label{sec:Algorithm_bs}
With Lemma \ref{lem:property_2}, by utilizing the monotonicity and continuity of $ R_{\alpha}(\tr(\textcolor{blue}{\bm \Sigma})) $ in $ \alpha $, an approximation of the unique value $ \alpha^{*} $ can be efficiently found by the binary search procedure in Algorithm \ref{alg}, where the parameter $\delta$ is a positive number close to 0 that controls the precision.

\begin{algorithm}[t]
\setstretch{1.15}
\normalsize
\caption{BinarySearch $( \textcolor{blue}{\bm \Sigma} , \delta )$}\label{alg:bisection}
\begin{algorithmic}[1]\label{alg}
\STATE $\alpha_{\rm L}\leftarrow 0, \alpha_{\rm R}\leftarrow 1$;
\STATE $ \alpha^{*}\leftarrow\frac{\alpha_{\rm L}+\alpha_{\rm R}}{2}$;
\WHILE {($ |R_{\alpha}(\tr(\textcolor{blue}{\bm \Sigma})) | > \delta $)}
\IF{$ R_{\alpha}(\tr(\textcolor{blue}{\bm \Sigma})) < 0 $}
\STATE $ \alpha_{\rm L} \leftarrow \alpha^{*} $;
\ELSE
\STATE $ \alpha_{\rm R} \leftarrow \alpha^{*} $;
\ENDIF
\STATE $\alpha^{*} \leftarrow \frac{\alpha_{\rm L} + \alpha_{\rm R}}{2}$;
\ENDWHILE
\RETURN $\alpha^{*} $;
\end{algorithmic}
\end{algorithm}

\subsection{Numerical Results}
In general, the distribution of real data might be degenerate \cite{chan2022redunet, ma2007segmentation}, for example, for image-type data, features (pixels) usually have correlations with each other. It leads to $ \textcolor{blue}{\bm \Sigma} $ being singular (the condition number $ \kappa $ being infinity), which is an ill-conditioned matrix. According to Corollary \ref{corollary:1}, the lower bound of $ \frac{R_{\alpha^{*}}(D)-R(D)}{n} $ always tends to negative infinity when $ \textcolor{blue}{\bm \Sigma} $ is singular. In this subsection, we investigate non-singular and singular matrices separately.

\subsubsection{Non-Singular Covariance Matrix}
\begin{figure}[ht]
\centering
\subfigure[]{
\begin{minipage}[b]{0.48\textwidth}
\label{fig:subfig_a}
\includegraphics[width=1\linewidth]{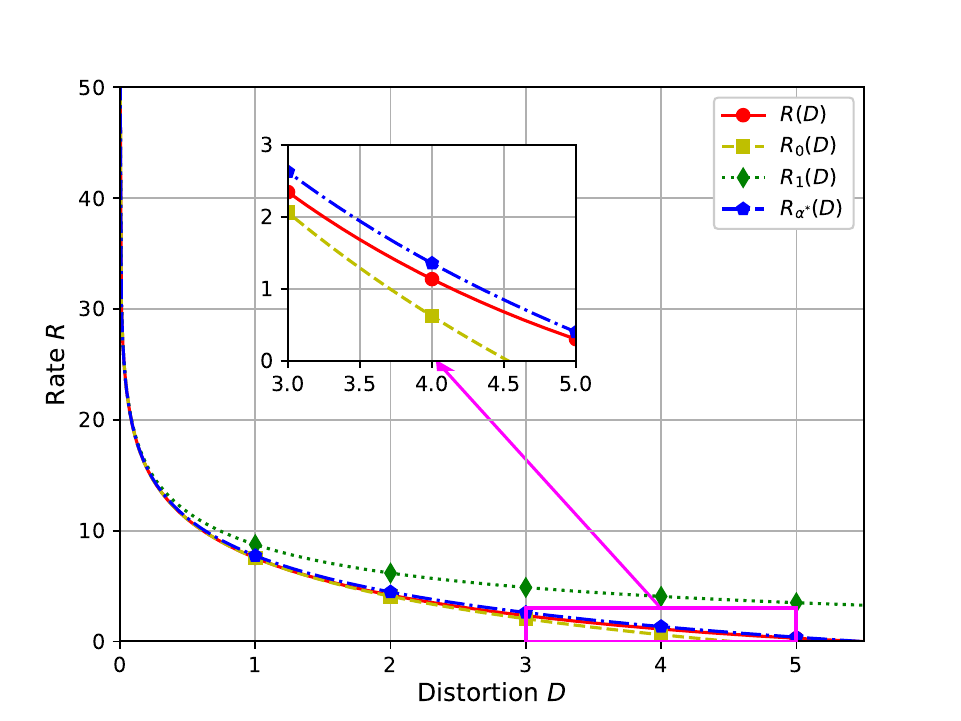}\vspace{4pt}
\end{minipage}}
\subfigure[]{
\begin{minipage}[b]{0.48\textwidth}
\label{fig:subfig_b}
\includegraphics[width=1\linewidth]{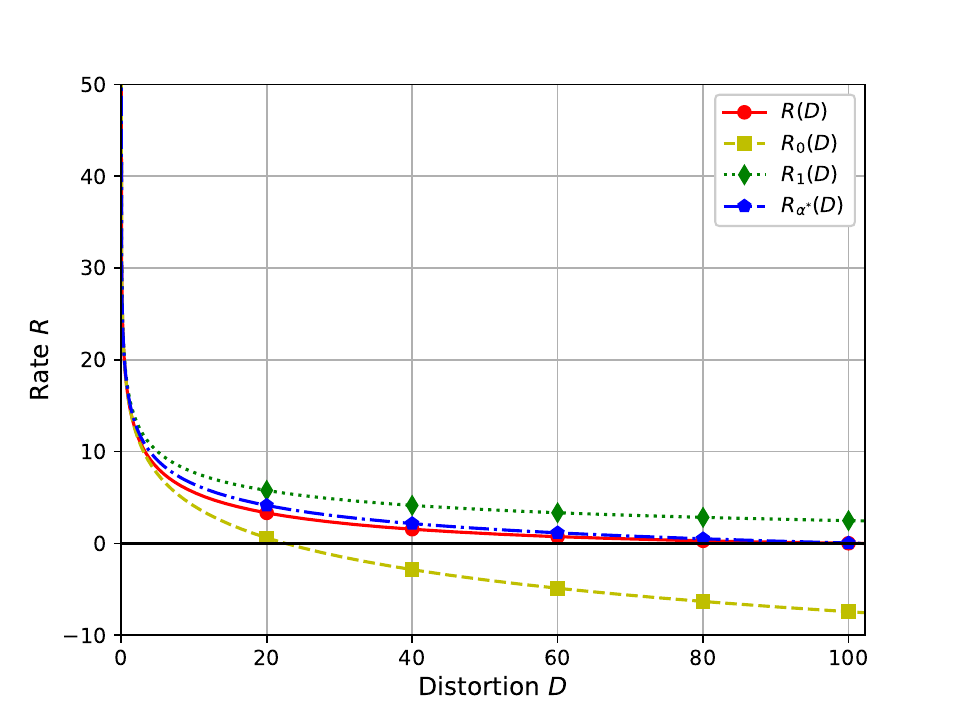}\vspace{4pt}
\end{minipage}}
\caption{Comparison of $R(D)$, $R_0(D)$, $R_1(D)$, and $R_{\alpha^*}(D)$: (a) $\mathbf{\Sigma}=\text{diag}(1.0, 0.9, 0.8, 0.7, 0.6, 0.5, 0.4, 0.3,$ $ 0.2, 0.1)$; (b) $\mathbf{\Sigma}=\text{diag}(51.2, 25.6, 12.8, 6.4, 3.2, 1.6,$ $ 0.8, 0.4, 0.2, 0.1)$.}
\label{fig:non-singular results}
\end{figure}
Let the matrix $\textcolor{blue}{ \bm \Sigma }$ be non-singular, for two $10$-dimensional Gaussian random vectors, the RD curves are illustrated in Fig. \ref{fig:non-singular results}. By \eqref{eq:RD} and \eqref{eq:RaDexpression}, both $R(D)$ and $R_{\alpha}(D)$ are determined by the eigenvalues of $\textcolor{blue}{\bm \Sigma}$. Without loss of generality, diagonal matrices $\textcolor{blue}{\bm \Sigma}$ are chosen. In particular, Fig. \ref{fig:subfig_a} illustrates the case when all the eigenvalues are similar with condition number $\kappa=10$ (which forms an arithmetic sequence, i.e., $\textcolor{blue}{\bm \Sigma}=\diag(1.0, 0.9, 0.8, 0.7, 0.6, 0.5, 0.4, 0.3, 0.2, 0.1)$); Fig. \ref{fig:subfig_b} illustrates a matrix with heterogeneous eigenvalues with a condition number $\kappa=512$ (which forms a geometric sequence, i.e., $\textcolor{blue}{\bm \Sigma}=\diag(51.2, 25.6, 12.8, 6.4, 3.2, 1.6, 0.8, 0.4, 0.2, 0.1)$). For comparison, we plot $R(D)$, $R_0(D)$, $R_1(D)$, and $R_{\alpha^*}(D)$ in Fig. \ref{fig:non-singular results}. The numerical results show that the function $ R_{\alpha^*}(D) $ provides a more accurate approximation of $ R(D) $ compared to $ R_{0}(D) $ and $ R_{1}(D) $, regardless of whether $ \kappa $ is large or small.

\begin{figure}[t]
\centering
\includegraphics[scale=0.48]{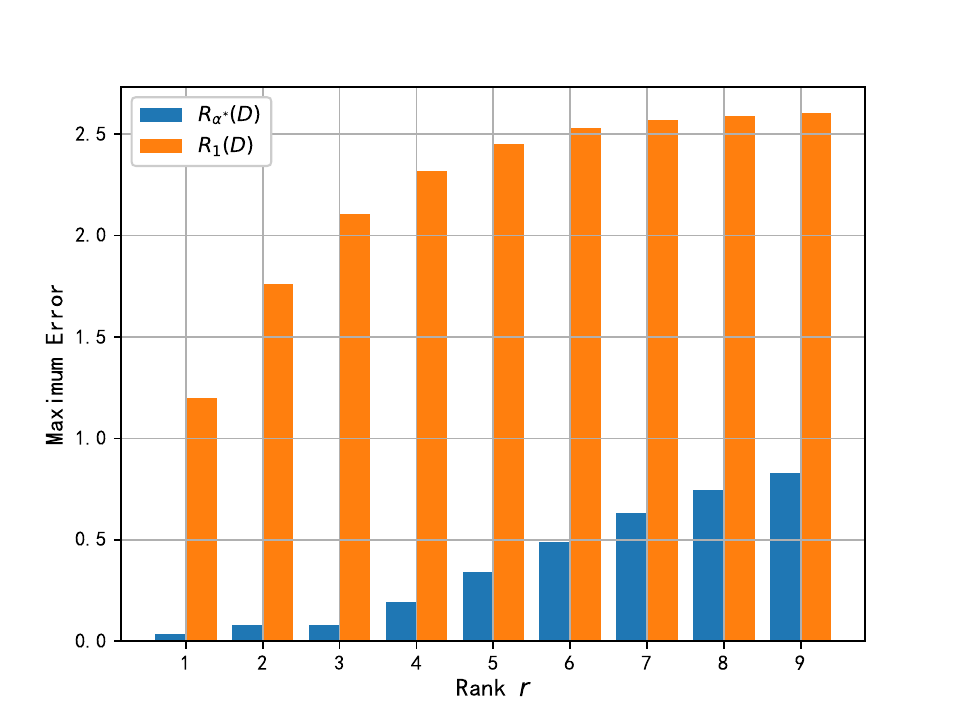}
\caption{Comparison of $ R_{1}(D) $ and $R_{\alpha^{*}}(D) $. The horizontal axis represents the rank $r $ of $ \mathbf{\Sigma} $, and the vertical axis represents the absolute value of the maximum error between the approximation function and $ R(D) $ over the interval $ D \in (0, \text{tr}(\mathbf{\Sigma)}] $.}
\label{fig:singular_subfig_a}
\end{figure}

\subsubsection{Singular Covariance Matrix}

\begin{figure}[ht]
\centering
\subfigure[]{
\begin{minipage}[b]{0.48\textwidth}
\label{fig:singular_subfig_b}
\includegraphics[width=1\linewidth]{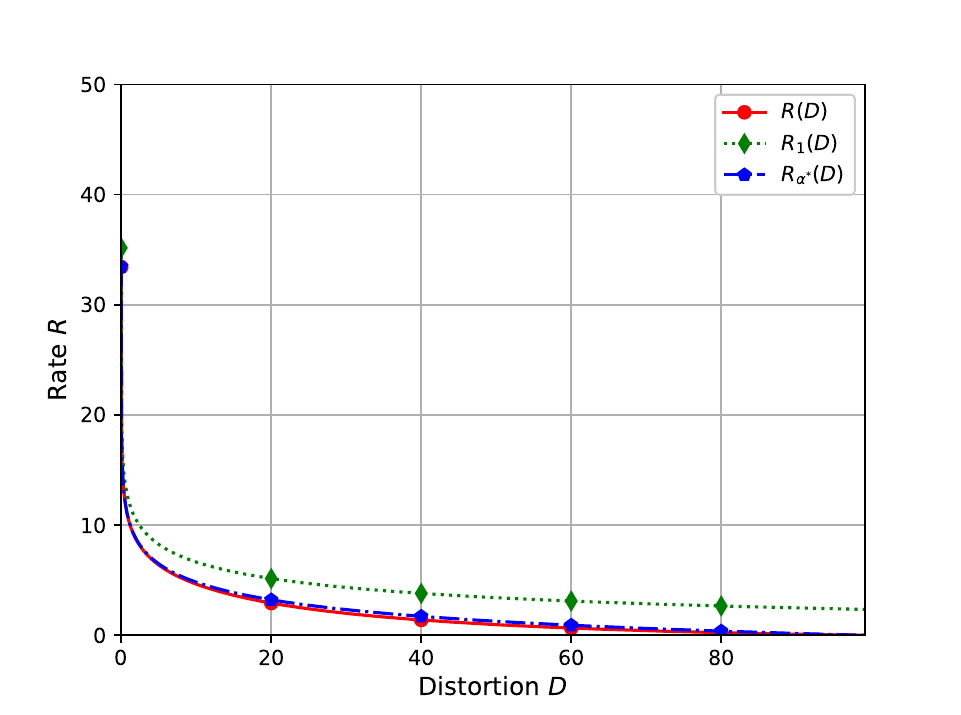}\vspace{4pt}
\end{minipage}}
\subfigure[]{
\begin{minipage}[b]{0.48\textwidth}
\label{fig:singular_subfig_c}
\includegraphics[width=1\linewidth]{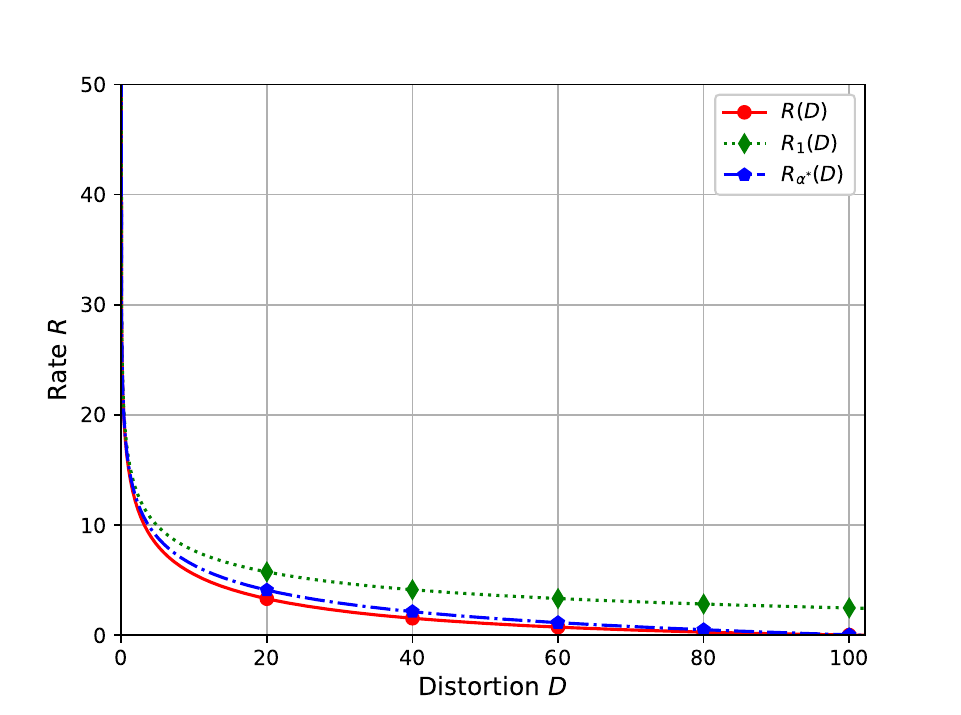}\vspace{4pt}
\end{minipage}}
\caption{Comparison of $R(D)$, $R_1(D)$, and $R_{\alpha^*}(D)$: (a) $ r=5 $, and $\mathbf{\Sigma}=\text{diag}(51.2, 25.6, 12.8, 6.4, 3.2,$ $  0, 0, 0, 0, 0)$; (b) $ r=9 $, and $\mathbf{\Sigma}=\text{diag}(51.2, 25.6, 12.8,$ $  6.4, 3.2, 1.6, 0.8, 0.4, 0.2, 0)$.}
\label{fig:singular results}
\end{figure}

Let $ \textcolor{blue}{\bm \Sigma} $ be a singular matrix with heterogeneous eigenvalues and $ r $ be the rank of the matrix $\textcolor{blue}{\bm \Sigma}$. We compare approximation function curves for various degenerate distributions (the rank of $ \textcolor{blue}{\bm \Sigma} $ decreases as the distribution becomes more degenerate). The covariance matrix $ \textcolor{blue}{\bm \Sigma}=\diag(\lambda_{1}, \lambda_{2}, \ldots, \lambda_{r}, 0, \ldots, 0) $, where $ \lambda_{i}=\frac{2^{10-i}}{10}$, $i=1,2, \ldots,r $ (similar to the scenario depicted in Fig. \ref{fig:subfig_b}, where the sequence of the non-zero eigenvalues is a geometric sequence). Since the function $ R_{0}(D) \rightarrow -\infty$ for any value of $ D $ when $ \textcolor{blue}{\bm \Sigma} $ is singular, the function $ R_{0}(D) $ is not compared with $ R_{1}(D) $ and $ R_{\alpha^{*}}(D) $. Fig. \ref{fig:singular_subfig_a} contrasts the approximation accuracy of $ R_{\alpha^{*}}(D) $ and $ R_{1}(D) $ for different $r$, which shows the difference between $\max_{D \in (0, \tr(\textcolor{blue}{\bm \Sigma})]} |R_{1}(D)-R(D)|$ and $\max_{D \in (0, \tr(\textcolor{blue}{\bm \Sigma})]}|R_{\alpha^{*}}(D)-R(D)| $. The numerical results indicate that $ R_{\alpha^{*}}(D) $ is more accurate than $ R_{1}(D) $ for any rank $ r \in [1, 9] $ of the matrix $ \textcolor{blue}{\bm \Sigma} $. As shown in Fig. \ref{fig:singular_subfig_a}, the difference between $ R_{\alpha^{*}}(D) $ and $ R_{1}(D) $ increases as $ r $ decreases, which shows that the greater the degradation, the greater the advantage of the new estimate. Moreover, as illustrated in Fig. \ref{fig:singular_subfig_b} and Fig. \ref{fig:singular_subfig_c}, the function $ R_{\alpha^{*}}(D) $ is still close to the exact RD function $ R(D) $, even though $ \textcolor{blue}{\bm \Sigma} $ is a singular matrix with heterogeneous eigenvalues.

In summary, whether $ \textcolor{blue}{\bm \Sigma} $ is singular or not, the function $ R_{\alpha^{*}}(D) $ is more accurate than $ R_{0}(D) $ and $ R_{1}(D) $, especially for smaller $ \kappa $. The numerical results support Corollary \ref{corollary:1}.

\subsection{Improving Approximation Accuracy through Data Pre-processing for High-Dimensional Data}  \label{Part_Data Pre-processing}

\begin{figure}[ht]
\centering
\includegraphics[scale=0.55]{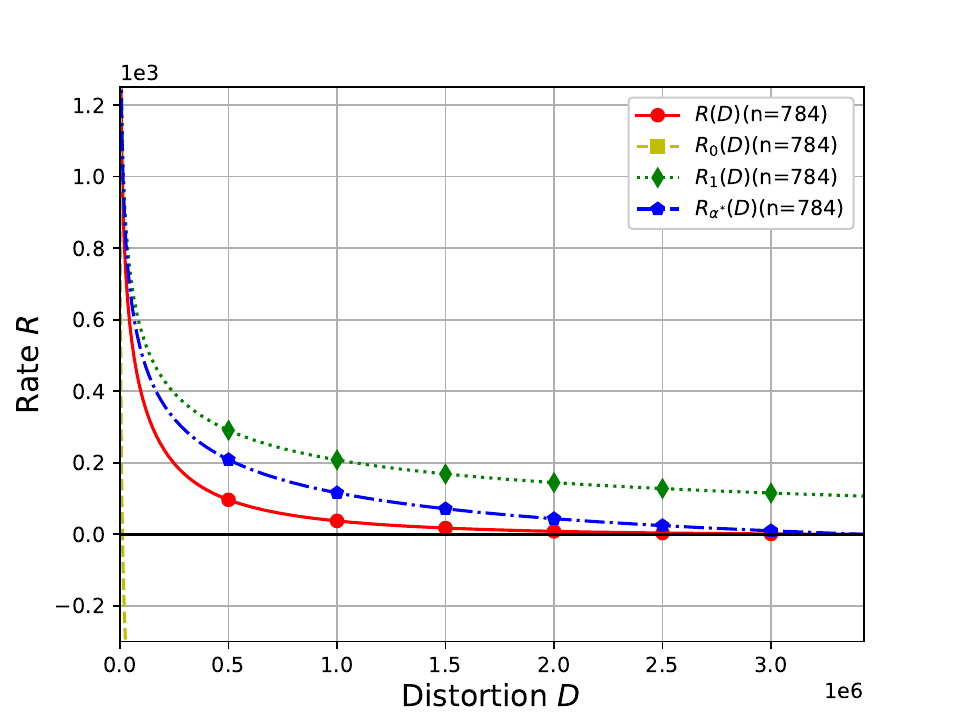}
\caption{Comparison of $R(D)$, $R_0(D)$, $R_1(D)$, and $R_{\alpha^*}(D)$ for MNIST training dataset, and the dimension $ n=784 $.}
\label{fig:data mnist}
\end{figure}

In the previous discussion, the function $ R_{\alpha^{*}}(D) $ is a good approximation when dimension $ n $ is small ($n=10$). However, Corollary \ref{corollary:1} indicates that as $ n $ increases, the upper and lower bounds of $ R_{\alpha^{*}}(D)-R(D) $ become relaxed. When the dimension $ n $ is large, e.g., the covariance matrix $\textcolor{blue}{\bm \Sigma} $ is estimated from the high-dimensional data (e.g. image-type data), the approximation accuracy of $ R_{\alpha^{*}}(D) $ might decrease significantly\footnote{For high-dimensional data, the unbiased estimate of covariance is not only high-dimensional, but also prone to be ill-conditioned, especially when the data dimension is larger than the sample size \cite{chen2000new}.}. Fig. \ref{fig:data mnist} shows that $ R_{\alpha^{*}}(D) $ is no longer accurate with large $ n $ and $ \kappa $. This motivates us to employ Principal Component Analysis (PCA), which is a common data pre-processing technique. PCA reduces the dimension of the data by applying a linear transformation to the original data, thus retaining the most critical features, i.e., the dimensions with large eigenvalues \cite{abdi2010principal}. Therefore, after using PCA reduces $ n $ and $ \kappa $, the approximation of $R(D)$ can be more accurate when applied to the pre-processed data.

Let $\textcolor{blue}{\bm Y}=[\bm y_1,\ldots,\bm y_m] \in \mathbb{R}^{p\times m}$ be the data matrix, where $\bm y_1,\ldots,\bm y_m$ are $p$-dimensional samples. The dimension is reduced by extracting the first $n$ principal components \cite{abdi2010principal}. The data matrix $\textcolor{blue}{\widetilde{\bm Y}} \in \mathbb{R}^{n \times m}$ after dimension reduction is
\begin{align}
\textcolor{blue}{\widetilde{\bm Y}=\widetilde{\bm U}^{\rm T}\bm{Y},}
\end{align}
where \textcolor{blue}{$\widetilde{\bm U}\in\mathbb{R}^{p\times n}$} is a matrix containing the top $ n $ principal components of the covariance of the data $ \textcolor{blue}{\bm Y} $ \cite{abdi2010principal}. The estimate of the covariance matrix is given by \textcolor{blue}{$\frac{1}{m-1}\widetilde{\bm Y}\widetilde{\bm Y}^{\rm T}$}. The RD function or its approximations are computed from the corresponding functions (e.g. $R(D)$, $R_0(D)$, $R_1(D)$, and $R_{\alpha^*}(D)$) based on the above covariance matrix estimation.

\begin{figure}[ht]
\centering
\subfigure[]{
\begin{minipage}[b]{0.48\textwidth}
\label{fig:data preprocessing_a}
\includegraphics[width=1\linewidth]{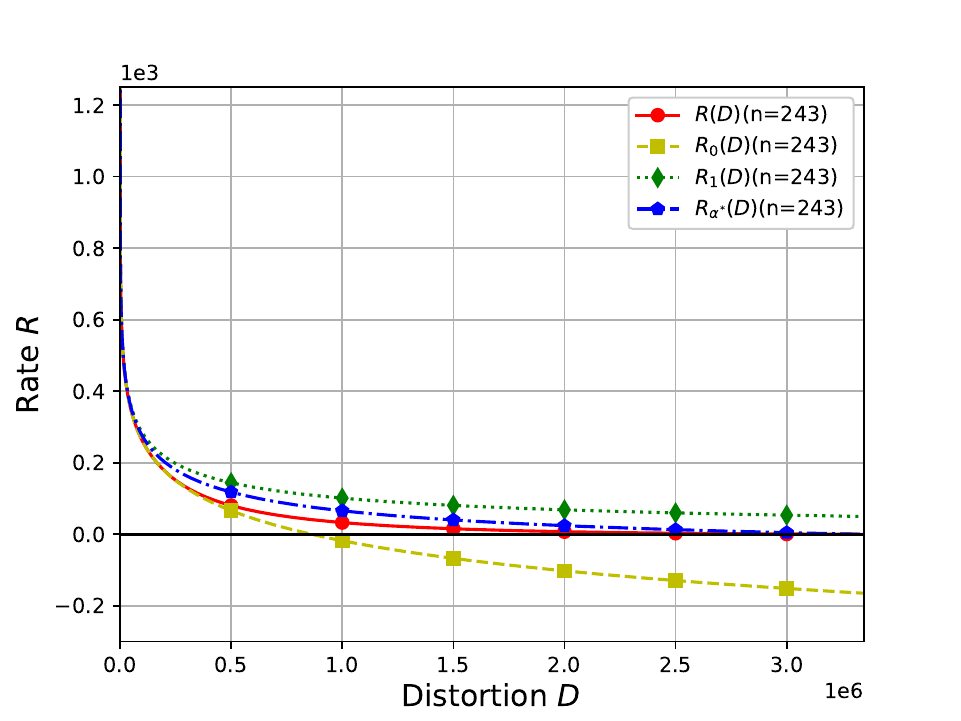}\vspace{4pt}
\end{minipage}}
\subfigure[]{
\begin{minipage}[b]{0.48\textwidth}
\label{fig:data preprocessing_b}
\includegraphics[width=1\linewidth]{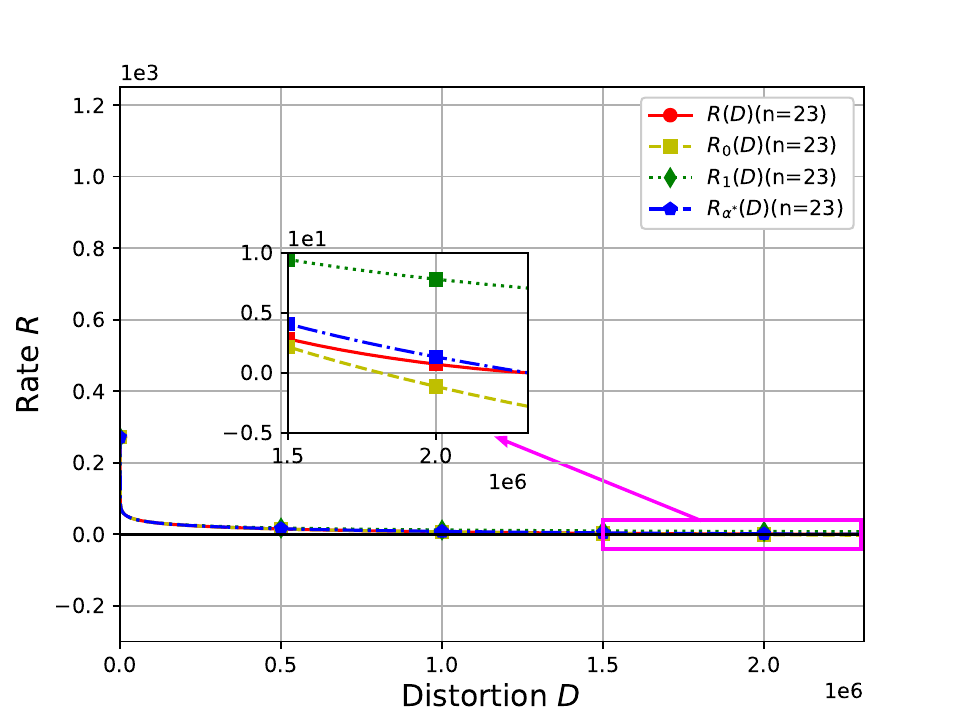}\vspace{4pt}
\end{minipage}}
\caption{Comparison of $R(D)$, $R_0(D)$, $R_1(D)$, and $R_{\alpha^*}(D)$ for MNIST training dataset after PCA pre-processing. (a) $ n=243 $; (b) $ n=23$.}
\label{fig:data preprocessing}
\end{figure}

To verify the effects of data pre-processing, we compare $R(D)$, $R_0(D)$, $R_1(D)$, and $R_{\alpha^*}(D)$ for a high-dimensional dataset, i.e., MNIST dataset, which contains $60000$ handwritten training samples of dimension $ p=28\times 28=784$ and $ 10000 $ testing samples. In Fig. \ref{fig:data preprocessing}, we plot two groups of curves using $ \textcolor{blue}{\bm \Sigma} $ obtained from the training dataset, which corresponds to $n=243$ and $23$, respectively. The unbiased estimate of the covariance of $ \textcolor{blue}{\bm \Sigma} $ is an ill-conditioned matrix when $ 243 $, while it is a non-ill-conditioned matrix when $ n=23 $. The condition numbers are approximately equal to $512$ and $10$, respectively\footnote{To contrast with Fig. \ref{fig:non-singular results}, we selected $ n=243 $ and $ 23 $, which have the condition numbers closest to $ 512 $ and $ 10 $, respectively.}.

Fig. \ref{fig:data preprocessing} indicates that the function $ R_{\alpha^{*}}(D) $ outperforms $ R_{0}(D) $ and $ R_{1}(D) $ for both ill-conditioned and non-ill-conditioned matrices, it can be observed that the accuracy of $ R_{\alpha^{*}}(D) $ increases as $ n $ and $ \kappa $ decreases. Compared with $n=784$, the maximum approximation errors of $ R_{\alpha^{*}}(D)$ for $n=243$ and $23$ are decreased by $69.88\%$ and $98.92\%$ respectively. This improvement comes from the fact that PCA reduces the dimension and condition number of the covariance matrix. Overall, PCA indeed improves the approximation accuracy of $ R_{\alpha^{*}}(D) $.

\section{Application: AR-ReduNet} \label{sec:applications}
Most neural network architectures and their components are typically designed based on years of trial and error and then deployed as a black-box. Recently, a white-box neural network for classification called ReduNet was proposed based on the Maximal Coding \textcolor{blue}{Rate} Reduction (MCR$^2$) principle, where the objective function was constructed using multivariate Gaussian RD functions \cite{chan2022redunet}. A popular assumption in a high-dimensional classification problem is that each class has a relatively low-dimensional intrinsic structure \cite{wright2022high}. ReduNet is designed to learn the low-dimensional intrinsic representations of high-dimensional data by extracting features, such that features of samples from different classes belong to different linear subspaces, while those from the same class are highly correlated in the sense that they belong to the same linear subspace \cite{yu2020learning, chan2022redunet}. This white-box neural network interprets modern deep networks from the perspective of data compression and discriminative representation \cite{chan2022redunet}, which play a crucial role in constructing neural networks with understandable structures. In this section, we derive a white-box neural network in the context of a classification problem by using the proposed approximation of the multivariate Gaussian RD function to replace that in ReduNet.

In ReduNet, the objective function is constructed using the RD function as building blocks, where $R_1(D)$ is used as an approximation for $R(D)$. In this paper, inspired by the parametric approximation, we use $R_{\alpha}(D)$ to approximate $R(D)$ in ReduNet, where $\alpha$ is updated in each iteration, according to the current estimate of the covariance matrix. The new network will be referred to as Adaptive Regularized ReduNet (AR-ReduNet).

\subsection{AR-ReduNet}
Let $\textcolor{blue}{\bm X_{\mathrm{data}}}=[\bm{x}_{1}, \ldots, \bm{x}_{m}] \in \mathbb{R}^{n \times m} $ be the data matrix of sample size $m$, where each $\bm x_i\in\mathbb{R}^n$ is the $i$-th $n$-dimensional sample, belonging to one of $k$ classes. For a given precision $ \epsilon>0 $, AR-ReduNet extracts the feature $\textcolor{blue}{\bm Z}=[\bm z_1,\ldots,\bm z_m]\in\mathbb{R}^{n\times m}$
by optimizing
\begin{align}
    \underset{\textcolor{blue}{\bm Z}}{\rm{maximize}} \quad \Delta \widetilde{R}(\textcolor{blue}{\bm Z},  \epsilon, \textcolor{blue}{\bm \Pi})&=\widetilde{R}^{\rm e}(\textcolor{blue}{\bm Z}, \epsilon)  \nonumber \\
    &-\sum_{j=1}^k\widetilde{R}^{\rm c}(\textcolor{blue}{\bm Z}, \epsilon | \textcolor{blue}{\bm \Pi_j}), \label{MCR2}\\
&{\rm{s. t.}}\quad \bm z_1,\ldots,\bm z_m\in\mathbb{S}^{n-1},\notag
\end{align}
where $\mathbb{S}^{n-1}$ denotes the unit ball in $n$-dimensional space. The notations in \eqref{MCR2} are explained as follows:
\begin{enumerate}
\item The membership of the samples is depicted by a set of $k$ diagonal matrices $\textcolor{blue}{\bm \Pi}=\{\textcolor{blue}{\bm \Pi_j}\}_{j=1}^k$, where $\textcolor{blue}{\bm \Pi_j}$  is the membership matrix of class $j$ defined by\footnote{By definition, the diagonal matrices $\textcolor{blue}{\bm \Pi}=\{\textcolor{blue}{\bm \Pi_j}\}_{j=1}^k$ lie in a simplex $\{\textcolor{blue}{\bm \Pi}:\pi_{ij}\geq 0, \sum_{j=1}^k\textcolor{blue}{\bm \Pi_j}=\textcolor{blue}{\bm I}\}$.}
\begin{align}
\textcolor{blue}{\bm \Pi_j}=\diag(\pi_{1,j}, \pi_{2,j}, \ldots, \pi_{m,j})\in \mathbb{R}^{m\times m},
\end{align}
with $\pi_{i,j}$ being the probability that the $i$-th sample belongs to class $j$. Notice that, in the training phase,
\begin{align}
\pi_{i,j}=\Bigg\{\begin{array}{ll}
1,&\mbox{if $\bm x_i$ is in class $j$}\\
0,&\mbox{else}  \label{Pi:training}
\end{array}.
\end{align}
\item The functions $\widetilde{R}^{\rm e}(\textcolor{blue}{\bm Z}, \epsilon)$ and $\widetilde{R}^{\rm c}(\textcolor{blue}{\bm Z}, \epsilon|\textcolor{blue}{\bm \Pi_j})$ are given by
\begin{subequations}\label{Rec}
\begin{align}
&\widetilde{R}^{\rm e}(\textcolor{blue}{\bm Z}, \epsilon)\triangleq  \nonumber\\
&\quad\quad\frac{1}{2}\log \det \Big(\alpha \textcolor{blue}{\bm I}+\frac{n}{m\epsilon^{2}}\textcolor{blue}{\bm Z \bm Z^{\rm T}}\Big),\label{Re}\\
&\widetilde{R}^{\rm c}(\textcolor{blue}{\bm Z}, \epsilon|\textcolor{blue}{\bm \Pi_j})\triangleq \nonumber\\
&\quad\quad\frac{\tr(\textcolor{blue}{\bm \Pi_j})}{2m}\log \det\left(\alpha_j \textcolor{blue}{\bm I}+\frac{n}{\tr(\textcolor{blue}{\bm \Pi_{j}})\epsilon^{2}}\textcolor{blue}{\bm Z \bm \Pi_{j} \bm Z^{\rm T}}\right),\label{Rc}
\end{align}
\end{subequations}
where the regularization parameters $\alpha$ and $\{\alpha_j\}_{j=1}^k$ are determined by\footnote{As the parameters $\alpha$ and $\{\alpha_j\}_{j=1}^k$ are determined by \eqref{implicit}, the notations of $\widetilde{R}^{\rm e}(\textcolor{blue}{\bm Z},\epsilon)$ and $\widetilde{R}^{\rm c}(\textcolor{blue}{\bm Z},\epsilon|\textcolor{blue}{\bm \Pi_j})$ do not include $\alpha$ or $\alpha_j$ as a subscript. }
\setlength{\abovedisplayskip}{0pt}
\setlength{\belowdisplayskip}{0pt}
\begin{subequations}\label{implicit}
\begin{align}
\frac{1}{2}\log \det \Big(\alpha \textcolor{blue}{\bm I}+\frac{n}{\tr(\textcolor{blue}{\bm Z \bm Z^{\rm T}})}\textcolor{blue}{\bm Z \bm Z^{\rm T}}\Big)&=0,\label{implicit:a}\\
\frac{1}{2}\log\det\Big(\alpha_j \textcolor{blue}{\bm I}+\frac{n}{\tr(\textcolor{blue}{\bm Z \bm \Pi_j \bm Z^{\rm T}})}\textcolor{blue}{\bm Z \bm \Pi_j \bm Z^{\rm T}}\Big)&=0. \label{implicit:b}
\end{align}
\end{subequations}
\end{enumerate}

Given the optimization problem \eqref{MCR2}, the feature $\textcolor{blue}{\bm Z}$ is obtained by deriving an $L$-layer network
\begin{align}\label{deepnet}
\textcolor{blue}{\bm Z^{(0)}}\rightarrow \textcolor{blue}{\bm Z^{(1)}}\rightarrow \ldots\rightarrow \textcolor{blue}{\bm Z^{(L)}},
\end{align}
where $\textcolor{blue}{\bm Z^{(\ell)}}=[\bm z_1^{(\ell)},\ldots,\bm z_{m}^{(\ell)}]\in\mathbb{R}^{n\times m}$ is the feature extracted in the $\ell$-th layer. The network is initialized by projecting the samples to $\mathbb{S}^{n-1}$, i.e.,
\begin{align}
\textcolor{blue}{\bm Z^{(0)}}\propto \textcolor{blue}{\bm X_{\mathrm{data}}}, \quad {\rm s.t.}~\bm z_1,\ldots,\bm z_m\in\mathbb{S}^{n-1}.
\end{align}
That is $\bm z_i^{(0)}=\frac{\bm x_{i}}{||\bm x_{i}||_2}$ for all $i=1,\ldots,m$. The feature $\textcolor{blue}{\bm Z^{(\ell+1)}}$ is obtained by gradient ascent at $\textcolor{blue}{\bm Z^{(\ell)}}$,  and the feature is projected to $\mathbb{S}^{n-1}$:
\begin{align}\label{Z:update}
 &\textcolor{blue}{\bm Z^{(\ell+1)}}\propto \textcolor{blue}{\bm Z^{(\ell)}}+\eta \cdot \frac{\partial \Delta \widetilde{R}}{\partial \textcolor{blue}{\bm Z}}\Big|_{\textcolor{blue}{\bm Z=\bm Z^{(\ell)}}}, \nonumber \\
 &\quad\quad\quad{\rm{s.t.}}\quad \bm  z_1^{(\ell+1)},\ldots, \bm z_m^{(\ell+1)}\in\mathbb{S}^{n-1},
\end{align}
where $\eta$ is the step size. 
  To be precise, define
\begin{align}
\textcolor{blue}{\bm E}&\triangleq\frac{n}{m \epsilon^{2}}\Big(\alpha \textcolor{blue}{\bm I}+\frac{n}{m\epsilon^{2}}\textcolor{blue}{\bm Z \bm Z^{\rm T}}\Big)^{-1} \in \mathbb{R}^{n \times n},\\
\textcolor{blue}{\bm C_j}&\triangleq\frac{ n}{m\epsilon^{2}}\Big(\alpha_j \textcolor{blue}{\bm I}+\frac{n}{\tr(\textcolor{blue}{\bm \Pi_j})\epsilon^{2}}\textcolor{blue}{\bm Z \bm \Pi_j \bm Z^{\rm T}}\Big)^{-1} \in \mathbb{R}^{n \times n},
\setlength{\abovedisplayskip}{0pt}
\setlength{\belowdisplayskip}{0pt}
\end{align}
the feature is updated as\footnote{Following the same steps in Ref. \cite{chan2022redunet}, it turns out that $\frac{\partial \widetilde{R}^{\rm e}(\textcolor{blue}{\bm Z},\epsilon)}{\partial \textcolor{blue}{\bm Z}}= \textcolor{blue}{\bm E \bm Z}$, and $\frac{\partial \widetilde{R}^{\rm c}(\textcolor{blue}{\bm Z},\epsilon|\textcolor{blue}{\bm \Pi_j})}{\partial \textcolor{blue}{\bm Z}}=\textcolor{blue}{\bm C_j \bm Z \bm \Pi_j}$ for $j=1,\ldots,k$. \vspace{-3pt}}
\begin{align}
&\textcolor{blue}{\bm Z^{(\ell+1)}}\propto \textcolor{blue}{\bm Z^{(\ell)}}+\eta \textcolor{blue}{\bm E^{(\ell)} \bm Z^{(\ell)}}-\eta\Big(\sum_{j=1}^k \textcolor{blue}{\bm C_{j}^{(\ell)} \bm Z^{(\ell)} \bm \Pi_j^{(\ell)}}\Big),\notag\\
&\quad\quad\quad\quad\quad {\rm{s.t.}}\quad \bm  z_1^{(\ell+1)},\ldots, \bm z_m^{(\ell+1)}\in\mathbb{S}^{n-1},
\end{align}
where $\textcolor{blue}{\bm E^{(\ell)}}$ and $\textcolor{blue}{\bm C_{j}^{(\ell)}}$  are the values of $\textcolor{blue}{\bm E}$ and $\textcolor{blue}{\bm C_j}$ evaluated at $\textcolor{blue}{\bm Z=\bm Z^{(\ell)}}$. The memberships $\textcolor{blue}{\bm \Pi_j^{(\ell)}}=\diag(\pi_{1,j}^{(\ell)},\ldots,\pi_{m,j}^{(\ell)})$ are determined by:
\begin{enumerate}
\item In the training phase, the labels of the samples are known, that is, $\textcolor{blue}{\bm \Pi_j^{(\ell)}=\bm \Pi_j}$, where $\textcolor{blue}{\bm \Pi_j}$ is determined by \eqref{Pi:training}.
\item In the test phase, $\textcolor{blue}{\bm \Pi_j^{(\ell)}}$ needs to be estimated. Following ReduNet \cite{chan2022redunet}, the softmax function is used on $\textcolor{blue}{\bm C_1^{(\ell)}\bm Z^{(\ell)},\ldots,\bm C_k^{(\ell)}\bm Z^{(\ell)}}$, i.e., $\textcolor{blue}{\bm \Pi_{j}^{(\ell)}}=\diag(\hat{\pi}_{1,j}^{(\ell)},\ldots,\hat{\pi}_{m,j}^{(\ell)})$, where $\hat\pi_{i,j}^{(\ell)}$ is specified by a hyperparameter $ \lambda $ that controls the uniformity\footnote{The membership  $\hat\pi_{i,j}^{(\ell)}$ can also be approximated in other ways, see Ref. \cite{chan2022redunet} for details. }:
\begin{align}
\hat{\pi}_{i,j}^{(\ell)}=\frac{\exp(-\lambda||\textcolor{blue}{\bm C^{(\ell)}_j}\bm z_i^{(\ell)}||_2)}{\sum_{j=1}^k\exp(-\lambda||\textcolor{blue}{\bm C^{(\ell)}_j}\bm z_i^{(\ell)}||_2)}.
\end{align}
\end{enumerate}

According to the MCR$^2$ principle \cite{yu2020learning}, for learned features to be discriminative, learned features from different classes
are preferred to be maximally incoherent to each other, while the ones from the same class should be highly correlated and coherent. Hence the coding rate of the whole set $\textcolor{blue}{\bm Z}$  should be as large as possible, while the ones of each class should be as small as possible. In \eqref{MCR2}, given a constraint that the mean square error does not exceed $\epsilon^{2}$, $\widetilde{R}^{\rm e}(\textcolor{blue}{\bm Z},\epsilon)$ depicts the minimum coding length of each feature when encoding all the features, while $\widetilde{R}^{\rm c}(\textcolor{blue}{\bm Z},\epsilon|\textcolor{blue}{\bm \Pi_j})$  depicts the weighted minimum coding length of each feature in class $j$ when encoding each class of the features separately.

The main distinction between ReduNet and AR-ReduNet is that, in order to approximate the coding length accurately, AR-ReduNet uses the parameterized forms in \eqref{Rec}, where the parameters $\alpha$ and $\{\alpha_j\}_{j=1}^k$ are updated in each layer according to \eqref{implicit}, which can be computed by Algorithm \ref{alg:bisection} with the covariance matrix $\textcolor{blue}{\bm \Sigma}$ evaluated by $\frac{1}{m}\textcolor{blue}{\bm Z \bm Z^{\rm T}}$ and $\frac{1}{\tr(\textcolor{blue}{\bm \Pi_j})}\textcolor{blue}{\bm Z \bm \Pi_j \bm Z^{\rm T}}$, respectively, while ReduNet adopts $ \alpha=1 $. To be clear, Algorithm \ref{alg:AR-ReduNet:1} and \ref{alg:AR-ReduNet:2} depict the training and testing of AR-ReduNet, respectively.

\begin{algorithm}[t]
\setstretch{1.15}
\caption{AR-ReduNet Training ($\textcolor{blue}{\bm X_{\mathrm{data}}, \bm \Pi}, \epsilon,\eta,\delta$)}\label{alg:AR-ReduNet:1}
\begin{algorithmic}[1]
\STATE  $ \textcolor{blue}{\bm Z^{(0)}}\leftarrow \Big[\frac{\bm x_1}{||\bm x_1||_2},\ldots,\frac{\bm x_m}{||\bm x_m||_2}\Big] $;  \quad // $\bm x_i$ is the $i$-th column of the data matrix $\textcolor{blue}{\bm X_{\mathrm{data}}}$.
\FOR{$\ell=0$ to $L-1$}
\STATE  $\alpha^{(\ell)}\leftarrow$ BinarySearch$( \textcolor{blue}{\bm Z^{(\ell)}(\bm Z^{(\ell)})^{\rm T}}/m , \delta )$;
\STATE  $ \textcolor{blue}{\bm E^{(\ell)}}\leftarrow\frac{n}{m\epsilon^{2}}\big(\alpha^{(\ell)} \textcolor{blue}{\bm I}+\frac{n}{m\epsilon^{2}} \textcolor{blue}{\bm Z^{(\ell)}(\bm Z^{(\ell)})^{\rm T}}\big)^{-1}$;
\FOR{$j=1$ to $k$}
\STATE $\alpha_j^{(\ell)}\leftarrow$BinarySearch$( \textcolor{blue}{\bm Z^{(\ell)} \bm \Pi_j(\bm Z^{(\ell)})^{\rm T}}/\tr(\textcolor{blue}{\bm \Pi_j}) , \delta )$;
\STATE $\textcolor{blue}{\bm C_{j}^{(\ell)}} \leftarrow \frac{n}{m\epsilon^{2}}\big(\alpha_j^{(\ell)} \textcolor{blue}{\bm I} + $ \\
$\quad\quad\quad\quad\quad\quad\quad \frac{n}{\tr(\textcolor{blue}{\bm \Pi_j})\epsilon^{2}} \textcolor{blue}{\bm Z^{(\ell)}\bm \Pi_j(\bm Z^{(\ell)})^{\rm T}}\big)^{-1} $;
\ENDFOR
\STATE\label{line:Zell}   $ \textcolor{blue}{\bm Z^{(\ell+1)}}\leftarrow \textcolor{blue}{\bm Z^{(\ell)}} + \eta \textcolor{blue}{\bm E^{(\ell)} \bm Z^{(\ell)}}-$ \\
$\quad\quad\quad\quad\quad\quad\quad\quad \eta\Big(\sum_{j=1}^k\textcolor{blue}{\bm C_{j}^{(\ell)}\bm Z^{(\ell)}\bm \Pi_j}\Big)$;
\STATE   $\textcolor{blue}{\bm Z^{(\ell+1)}}\leftarrow \Big[\frac{\bm z_{1}^{(\ell+1)}}{||\bm z_{1}^{(\ell+1)}||_2},\ldots,\frac{\bm z_{m}^{(\ell+1)}}{||\bm z_{m}^{(\ell+1)}||_2}\Big]$;\quad // $\bm z_i^{(\ell+1)}$ is the $i$-th column of $\textcolor{blue}{\bm Z^{(\ell+1)}}$ in Line \ref{line:Zell}.
\ENDFOR
\RETURN $ \textcolor{blue}{\bm Z^{(L)}, \{\bm E^{(\ell)}, \bm C_{1}^{(\ell)}, \bm C_{2}^{(\ell)}, \ldots, \bm C_{k}^{(\ell)}\}_{\ell=0}^{L-1}}$;
\end{algorithmic}
\end{algorithm}

\begin{algorithm}[t]
\setstretch{1.15}
\caption{AR-ReduNet Testing ($ \textcolor{blue}{\widetilde{\bm X}},$ $\textcolor{blue}{\{\bm E^{(\ell)}},$ $ \textcolor{blue}{\bm C_{1}^{(\ell)}},$  $\ldots,$ $ \textcolor{blue}{\bm C_{k}^{(\ell)}\}_{\ell=0}^{L-1}}, \eta, \lambda $)}\label{alg:AR-ReduNet:2}
\begin{algorithmic}[1]
\STATE  $ \textcolor{blue}{\widetilde{\bm Z}^{(0)}}\leftarrow \Big[\frac{\widetilde{\bm x}_1}{||\widetilde{\bm x}_1||_2},\ldots,\frac{\widetilde{\bm x}_m}{||\widetilde{\bm x}_m||_2}\Big]$;\quad // $\widetilde{\bm x}_i$ is the $i$-th column of data matrix $\textcolor{blue}{\widetilde{\bm X}}$.
\FOR{$\ell=0$ to $L-1$}
\FOR{$j=1$ to $k$}
\FOR{$i=1$ to $m$}
\STATE  $ \hat{\pi}_{i,j}^{(\ell)}\leftarrow \frac{\exp(-\lambda||\textcolor{blue}{\bm C^{(\ell)}_j}\bm{\widetilde{z}}_i^{(\ell)}||)}{\sum_{j=1}^k\exp(-\lambda||\textcolor{blue}{\bm C^{(\ell)}_j}\bm{\widetilde{z}}_i^{(\ell)}||)} $;
\ENDFOR
\STATE $\textcolor{blue}{\bm \Pi_j^{(\ell)}}\leftarrow\diag(\hat{\pi}_{1,j}^{(\ell)},\ldots,\hat{\pi}_{m,j}^{(\ell)})$;
\ENDFOR
\STATE\label{line:Zelltilde}     $ \textcolor{blue}{\widetilde{\bm Z}^{(\ell+1)}} \leftarrow\textcolor{blue}{\widetilde{\bm Z}^{(\ell)}} + \eta \textcolor{blue}{\bm E^{(\ell)} \widetilde{\bm Z}^{(\ell)}}-$ \\
$ \quad\quad\quad\quad\quad\quad\quad\quad \eta\Big(\sum_{j=1}^k\textcolor{blue}{\bm C_{j}^{(\ell)}\widetilde{\bm Z}^{(\ell)}\bm \Pi_j^{(\ell)}}\Big)$;
\STATE $\textcolor{blue}{\widetilde{\bm Z}^{(\ell+1)}}\leftarrow \Big[\frac{\widetilde{\bm z}_{1}^{(\ell+1)}}{||\widetilde{\bm z}_{1}^{(\ell+1)}||_2},\ldots,\frac{\widetilde{\bm z}_{m}^{(\ell+1)}}{||\widetilde{\bm z}_{m}^{(\ell+1)}||_2}\Big]$;\quad // $\widetilde{\bm z}_i^{(\ell+1)}$ is the $i$-th column of $\textcolor{blue}{\widetilde{\bm Z}^{(\ell+1)}}$ in Line \ref{line:Zelltilde}.
\ENDFOR
\RETURN $ \textcolor{blue}{\widetilde{\bm Z}^{(L)}} $.
\end{algorithmic}
\end{algorithm}

The features extracted in Algorithm \ref{alg:AR-ReduNet:2} can be used to determine which class the samples belong to. This needs a suitable classifier, which can be Support Vector Machine (SVM), $k$-Nearest-Neighbor ($k$NN), or Nearest Subspace (NS), etc. Among various classifiers, the NS classifier makes decisions based on the residuals of the features projected on the subspace, which spanned by the left singular vectors of the training set of each class \cite{tong2022incremental}. As AR-ReduNet and ReduNet try to separate  different types into distinct subspaces, the NS classifier is a coherent classifier to AR-ReduNet and ReduNet. Following ReduNet \cite{chan2022redunet}, in the following simulations, we choose the NS classifier. That is, the $i$-th testing sample $\widetilde{\bm x}_i$ is assigned to the class $\widehat{j}_i$ if
\begin{align}
\widehat{j}_i=\underset{t \in \{1, \ldots, k\}}{\arg \min}|| (\textcolor{blue}{\bm I-\bm U_{t} \bm U_{t}^{\rm T}})\widetilde{\bm z}_{i}^{(L)} ||_{2}^{2},
\end{align}
where $\textcolor{blue}{\bm U_t} \in \mathbb{R}^{n \times n}$ is the orthogonal matrix composed of left singular vectors of $\textcolor{blue}{\bm Z_j}$, the extracted features of training data in class $j$ by Algorithm \ref{alg:AR-ReduNet:1}.

\subsection{Simulation Results}
In our simulations, we compare the performance of ReduNet and AR-ReduNet, and evaluate the impact of PCA and parameter $ \epsilon $ (distortion $ D = \epsilon^{2} $ in $ R_{1}(D) $ and $ R_{\alpha^{*}}(D) $). The dataset used is MNIST, which contains a training set of 60,000 samples and a test set of 10,000 samples.

\subsubsection{The Accuracy of ReduNet and AR-ReduNet Based on PCA}

According to the numerical results in Section \ref{Part_Data Pre-processing}, PCA significantly reduces the inaccuracy of the approximation function, so it is reasonable to believe that PCA can improve the performance of white-box neural networks. We pre-process the dataset using PCA with different ratios of cumulative variance, and compare the classification accuracy of ReduNet and AR-ReduNet. The cumulative variance ratio $ P = \frac{C_{1}}{C_{2}} $ characterizes the number of principal components based on a percentage\footnote{For example, if we retain about $ 98\% $ of the cumulative variance, then $ n=261 $, because the sum of the top $ 261 $ eigenvalues is approximately equal to $98\%$ of the total variation.}, where $ C_{1} $ is the sum of the top $ n $ eigenvalues, and $ C_{2}=\tr(\textcolor{blue}{\bm \Sigma}) $. It is a commonly used measure in PCA \cite{jolliffe2002choosing}. The parameters of ReduNet and AR-ReduNet include: step size $\eta=0.5$, hyperparameter that controls the uniformity $\lambda=500$, precision of the binary search $\delta=10^{-8}$, number of layers $ L=1000 $. We conduct three sets of experiments, where $ \epsilon^{2}=0.3 $, $ 0.5 $, and $ 0.7 $.

The trends of the approximation error and classification accuracy with cumulative variance ratio are illustrated in Fig. \ref{fig:PCA_results}. It can be observed that
\begin{itemize}
  \item In Fig. \ref{fig:PCA_subfig_a}, for each set of experiments, PCA pre-processing indeed reduces the approximation error. However, as $ P $ decreases uniformly, the trend of the approximation error tends to be stable.
  \item In Fig. \ref{fig:PCA_subfig_b}, for the same $ \epsilon $, AR-ReduNet outperforms ReduNet due to more accurate approximation function. In addition, the simulation results indicate that there is optimal $ P $ that achieves the highest accuracy for each $ \epsilon $. The reason is that when the accuracy of the approximation functions approaches a plateau, as $ n $ is further reduced, the classification accuracy may decrease due to the loss of feature information.
\end{itemize}

On the other hand, since the inverse matrix of the covariance matrix has to be computed according to Algorithm \ref{alg:AR-ReduNet:1}, the time complexity of Algorithm \ref{alg:AR-ReduNet:1} is $ O((k+1)Ln^{3}) $. This results in a high computational cost for AR-ReduNet training when $ n $ is large (the same applies to ReduNet). Therefore, it is necessary to use PCA. Choosing the appropriate dimension $ n $ not only improves the classification accuracy of the white-box neural network, but also reduces the computational cost.

\subsubsection{The Effect of Precision $ \epsilon $ on ReduNet and AR-ReduNet}   \label{sec:Precision}

\begin{figure}[!ht]
\centering
\subfigure[]{
\begin{minipage}[b]{0.48\textwidth}
\label{fig:PCA_subfig_a}
\includegraphics[width=1\linewidth]{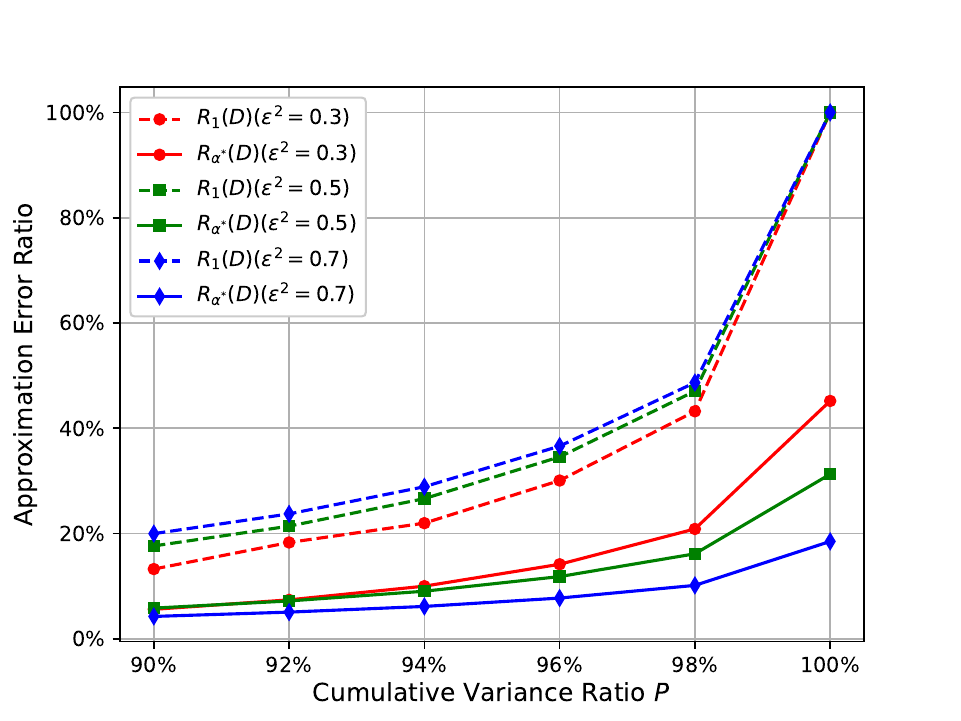}\vspace{4pt}
\end{minipage}}
\subfigure[]{
\begin{minipage}[b]{0.48\textwidth}
\label{fig:PCA_subfig_b}
\includegraphics[width=1\linewidth]{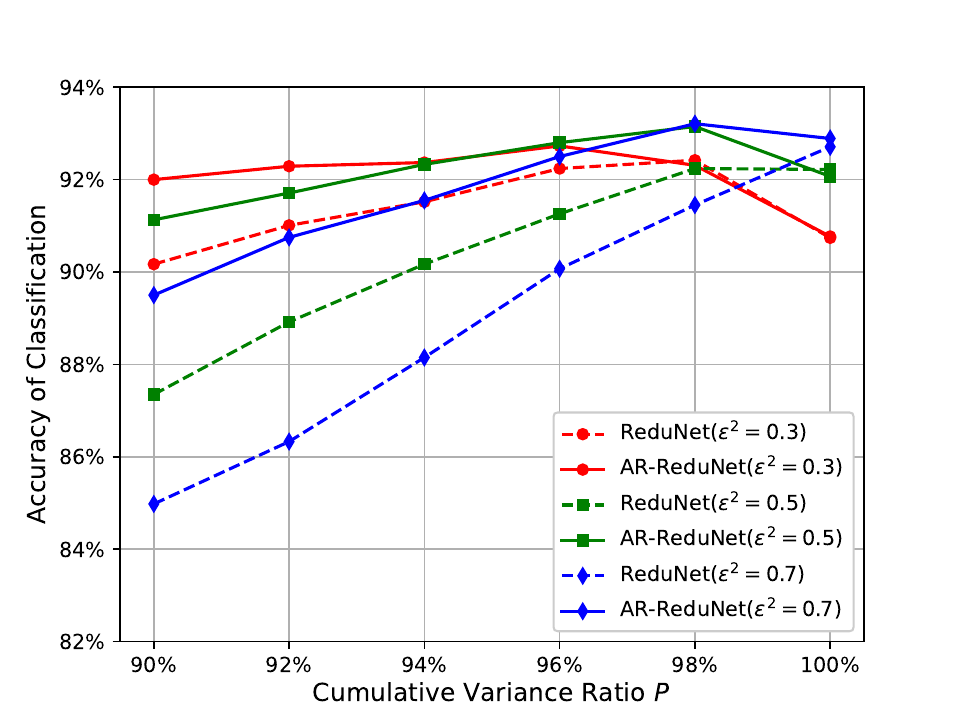}\vspace{4pt}
\end{minipage}}
\caption{Three sets of simulation results for $ \epsilon^{2}=0.3 $, $ 0.5 $ and $ 0.7 $. (a) The trends of the approximation error ratio of $ R_{1}(D) $ and $ R_{\alpha^{*}}(D) $ with the cumulative variance ratio. The approximation error ratio is equal to $ E_{\textrm{new}}/E_{\textrm{old}} $, where $E_{\textrm{new}}$ is the error of the approximation function with different $ P $, and $E_{\textrm{old}}$ is the error of $ R_{1}(D) $ at $ P=100\% $; (b) Comparison of the classification accuracy between ReduNet and AR-ReduNet.}
\label{fig:PCA_results}
\end{figure}

\begin{figure}[!ht]
\centering
\subfigure[]{
\begin{minipage}[b]{0.48\textwidth}
\label{fig:subfig_classification_a}
\includegraphics[width=1\linewidth]{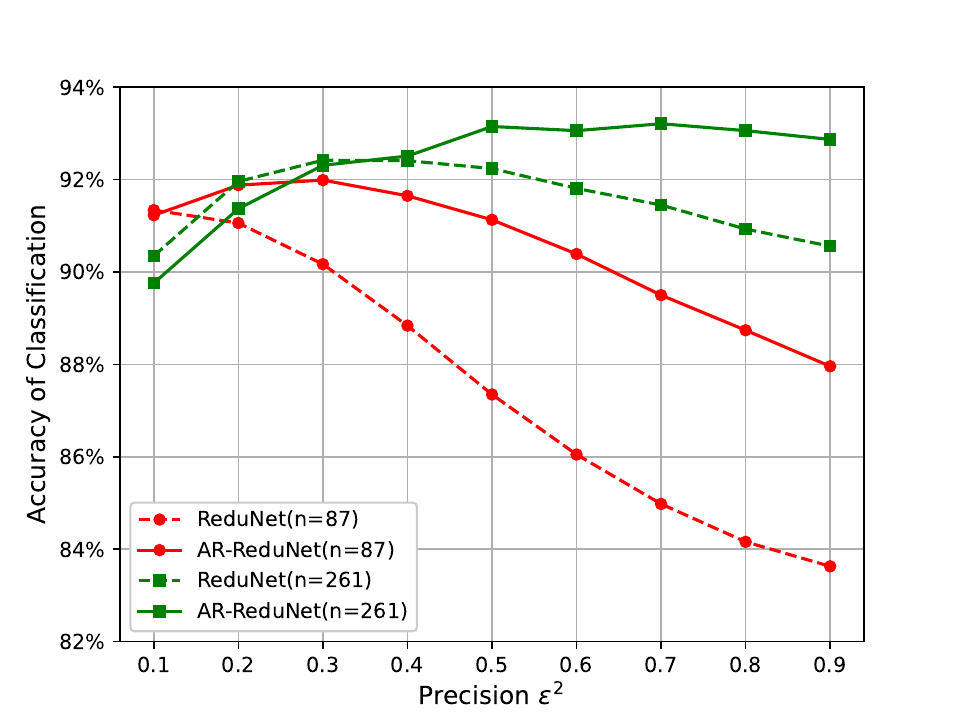}\vspace{4pt}
\end{minipage}}
\subfigure[]{
\begin{minipage}[b]{0.48\textwidth}
\label{fig:subfig_classification_b}
\includegraphics[width=1\linewidth]{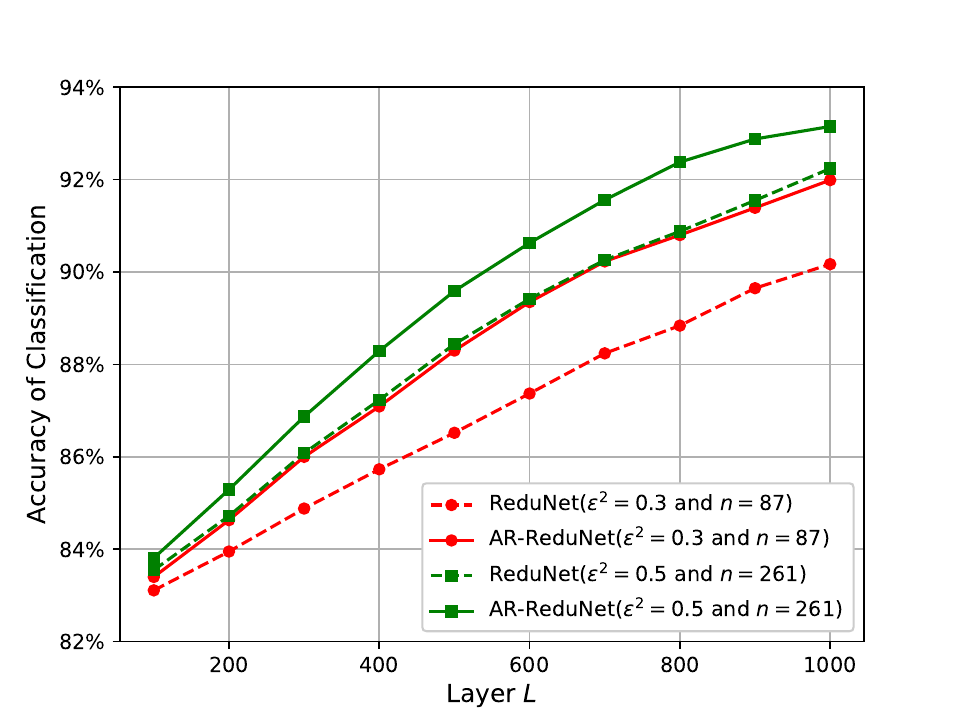}\vspace{4pt}
\end{minipage}}
\caption{The trends of classification accuracy of ReduNet and AR-ReduNet with (a) precision $\epsilon$ for given dimension $ n $; (b) the number of layers $ L $ for given $\epsilon$ and $n$.}
\label{fig:classification}
\end{figure}

In Ref. \cite{chan2022redunet}, ablation study shows that $ \epsilon $ plays an important role in the generalization performance of ReduNet, it is very usual to keep $ \epsilon $ as a free parameter that can be set by users \cite{lu2023interpretable}. Fig. \ref{fig:PCA_subfig_b} indicates that for each $ P $, selecting the appropriate parameter $ \epsilon $ can effectively improve the classification accuracy. Here we evaluate the effect of $ \epsilon $ on ReduNet and AR-ReduNet. The parameters of ReduNet and AR-ReduNet are specified as: $\eta=0.5$, $\lambda=500$, $\delta=10^{-8}$, $ L=1000 $, and $ \epsilon^{2} \in (0, 1) $. For the values of $ \kappa $, we perform two sets of experiments with $ \kappa \approx 546.57 $ ($P=98\%$) and $ 74.56 $ ($ P=90\% $). They represent the cases that the unbiased estimates of the covariance are ill-conditioned and non-ill-conditioned, respectively.

The classification accuracy trends of ReduNet and AR-ReduNet with precision $ \epsilon $ are illustrated in Fig. \ref{fig:subfig_classification_a}. It can be observed that AR-ReduNet outperforms ReduNet regardless of whether the unbiased estimate of the covariance is ill-conditioned or not. For example, for $ n=87 $, AR-ReduNet with $ \epsilon^{2}=0.3 $ outperforms ReduNet for any value of $ \epsilon^{2} $, while for $ n=261 $, AR-ReduNet with $ \epsilon^{2}=0.5 $ is also better than ReduNet for any value of $ \epsilon^{2} $. In addition, it can be observed that when the covariance matrix is ill-conditioned, choosing a large $ \epsilon $ can improve the classification accuracy, while when the covariance matrix is well-conditioned, a small $ \epsilon $ should be selected. Fig. \ref{fig:subfig_classification_b} illustrates the trends of the accuracy with the number of layers for fixed dimension and distortion pair, where we choose $ \epsilon^{2}=0.3 $ and $ 0.5 $ for dimension $ n=87 $ and $ 261 $, respectively, since they achieve extremely high accuracy for AR-ReduNet in Fig. \ref{fig:subfig_classification_a}. Fig. \ref{fig:subfig_classification_b} indicates that AR-ReduNet achieves a significant gain in accuracy than ReduNet due to function $R_{\alpha^{*}}(D)$.

\subsubsection{The Feature Extracted by ReduNet and AR-ReduNet}

\begin{figure}[!ht]
\centering
\subfigure[]{
\begin{minipage}[b]{0.48\textwidth}
\label{fig:subfig_feature_a}
\includegraphics[width=1\linewidth]{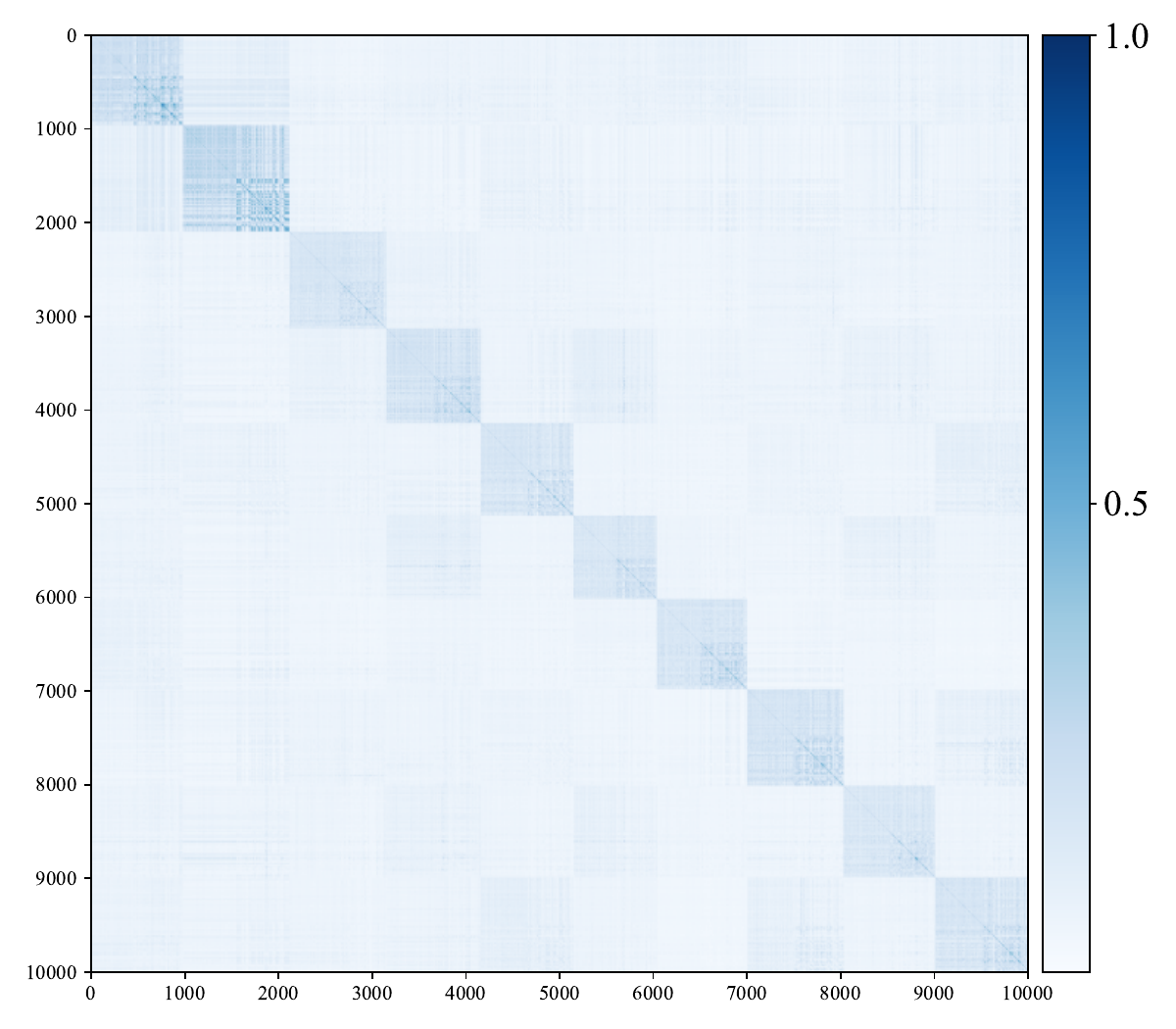}\vspace{4pt}
\end{minipage}}
\subfigure[]{
\begin{minipage}[b]{0.48\textwidth}
\label{fig:subfig_feature_b}
\includegraphics[width=1\linewidth]{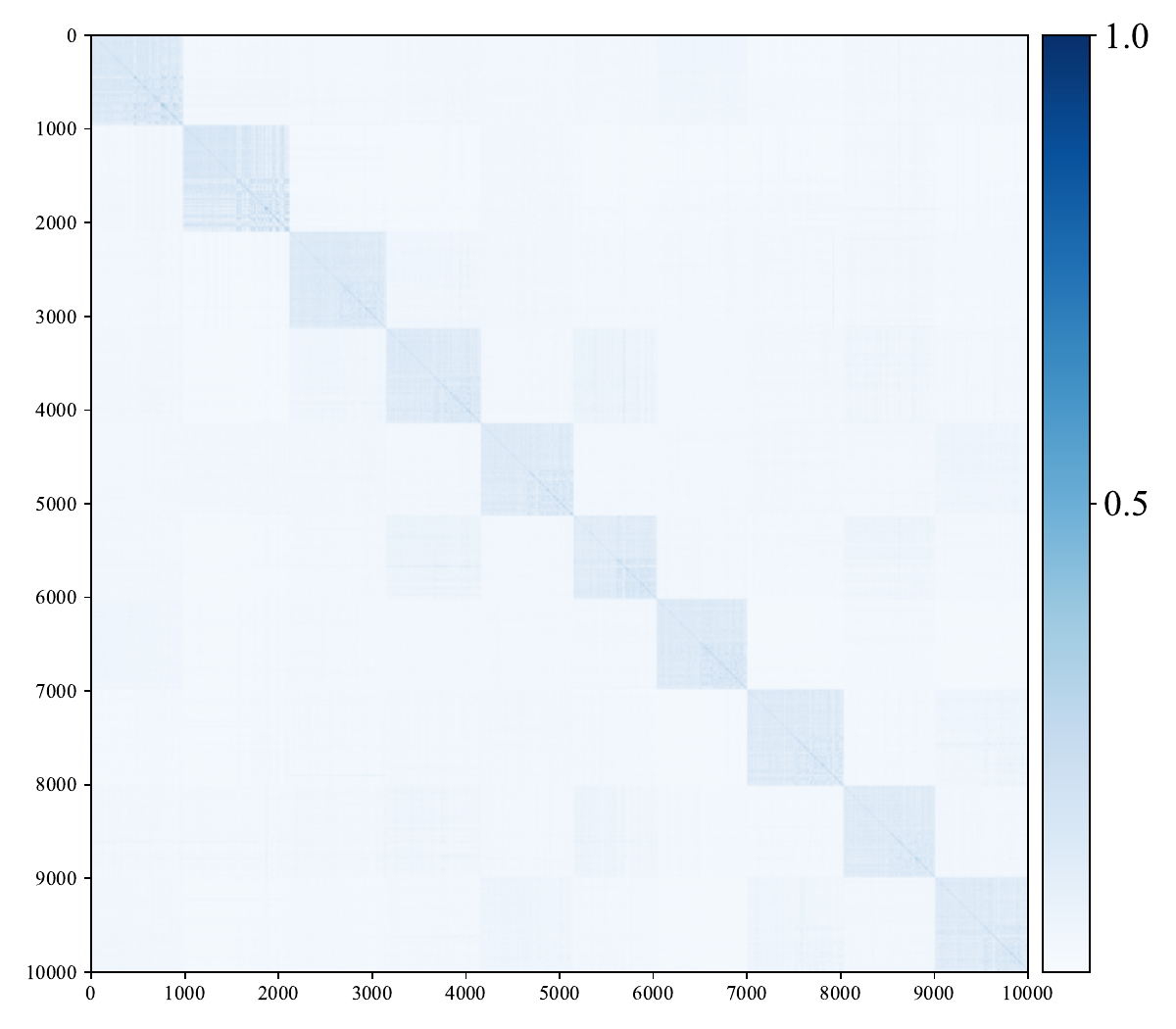}\vspace{4pt}
\end{minipage}}
\caption{The heatmaps of cosine similarities among testing samples. (a) the cosine similarities between features extracted by ReduNet; (b) the cosine similarities between features extracted by AR-ReduNet.}
\label{fig:feature}
\end{figure}

In Ref. \cite{chan2022redunet}, to evaluate the learned features, they compared the cosine similarities among the learned features extracted by ReduNet. The extracted features exhibit high correlations between features of the same class and low correlations between different classes. In Fig. \ref{fig:feature}, the cosine similarities between features extracted from 10000 testing samples by ReduNet and AR-ReduNet are illustrated. The parameters of ReduNet and AR-ReduNet are consistent with Section \ref{sec:Precision}, and fixed $n=261$ and $ \epsilon^{2}=0.5 $.

In Fig. \ref{fig:feature}, the intensity of the colors in the figure corresponds to the magnitude of the correlation, where darker colors indicate higher correlations and lighter colors indicate lower correlations. As can be seen from Fig. \ref{fig:feature}, the correlation of features extracted by AR-ReduNet is lower than that of ReduNet, e.g., the correlation between handwritten digits 0 and 1, 4 and 9, as well as 7 and 9.

In summary, AR-ReduNet achieves higher accuracy than ReduNet, especially when the condition number $ \kappa $ is small. In addition, for the same precision $ \epsilon $, AR-ReduNet optimizes network parameters more efficiently than ReduNet, due to the improved approximation for the RD function. The simulation results support the priority of using AR-ReduNet.

\section{Conclusions} \label{sec:conclusions}
Both theoretical and numerical results indicate that the proposed approximation $R_{\alpha^*}(D)$ is a rather accurate approximation of the multivariate Gaussian RD function $R(D)$, especially for the distributions with well-conditioned covariance matrices. Moreover, an application in a neural network for classification (AR-ReduNet) is derived by incorporating the approximation into ReduNet and adaptively updating the parameter $\alpha$ with the bisection algorithm. Simulation results support that improved approximation achieves higher classification accuracy and more efficient optimization.

Remarkably, the proposed approximation $R_{\alpha^*}(D)$ not only has a rather simple form and good analytical properties, but also has accurate numerical results for non-singular covariance matrices. It is expected that in addition to ReduNet, this approximation has more potential applications in other scenarios where the multivariate Gaussian RD function plays a central role, such as closed-loop transcription \cite{dai2022ctrl} and white-box transformers \cite{yu2023white}. In the future, it is worth investigating these models and improving the accuracy and optimization efficiency by using the approximation function.

\zihao{5-}
\noindent
\renewcommand\refname{\zihao{5}\textbf{References}}
\zihao{5-} \addtolength{\itemsep}{-1em}

\begin{biography}[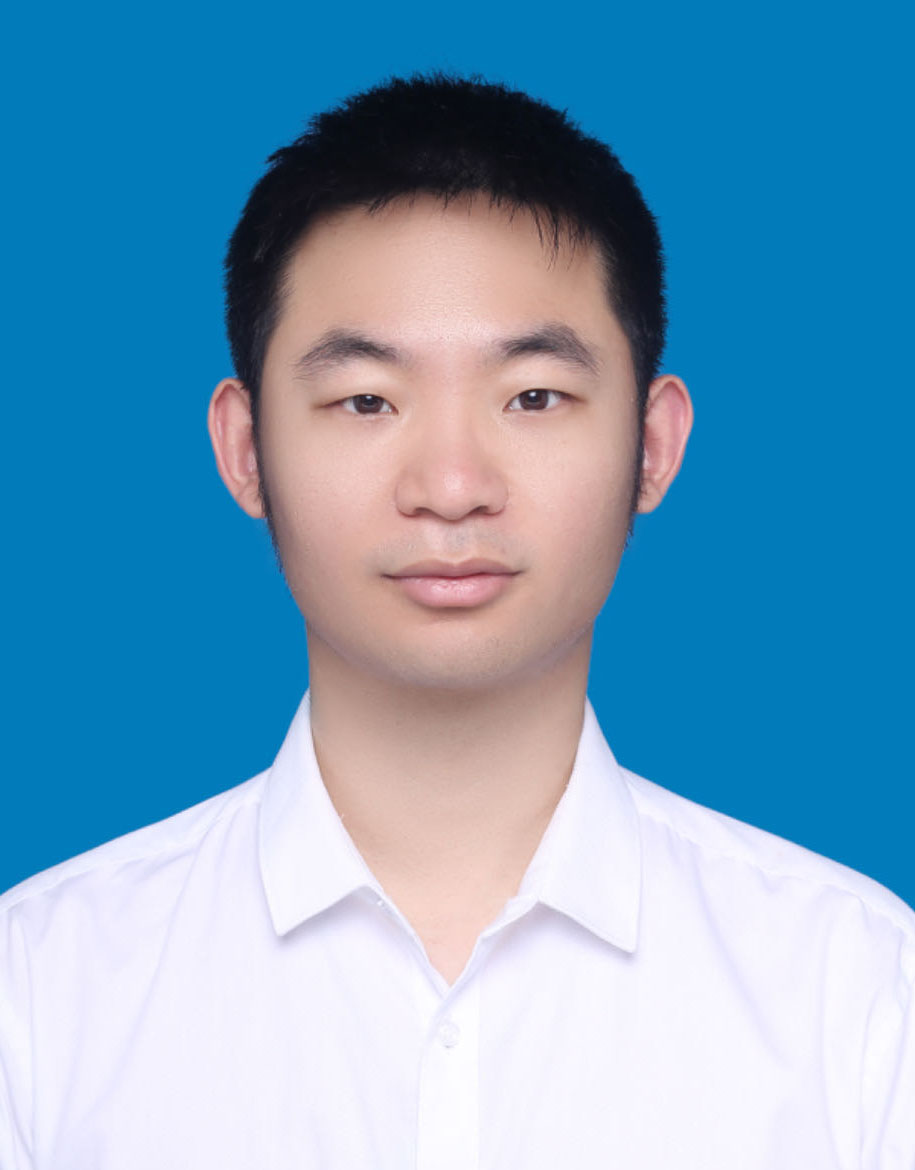]
\noindent
\textbf{Zhenglin Huang} received the B.S. degree in applied software engineering from Southwest Jiaotong University, Chengdu, China, in 2018 and the M.S. degree in applied data science and analysis from Brunel University London, in 2019. Currently, he is pursuing his Ph.D. degree in traffic information engineering and control at Southwest Jiaotong University, Chengdu, China. His research interest is the intersection of information theory and artificial intelligence.
\end{biography}

\begin{biography}[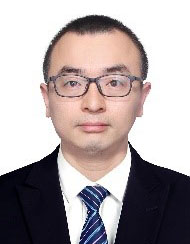]
\noindent
\textbf{Qifa Yan} received the B.S. degree in mathematics and applied mathematics from Shanxi University, Taiyuan, China, in 2010, and the Ph.D. degree in communication and information system from the School of Information Science and Technology, Southwest Jiaotong University, Chengdu, China, in 2017. From November 2017 to October 2019, he was a Joint Post-Doctoral Researcher with the T\'{e}l\'{e}com Paris, Institut Politechnique de Paris, and CentraleSup\'{e}lec, Paris-Saclay University, France. From January 2020 to January 2021, he was a Post-Doctoral Research Fellow with the Department of Electrical and Computer Engineering, University of Illinois Chicago, USA. He is currently an Associate Professor with the School of Information Science and Technology, Southwest Jiaotong University. His research interests include caching networks, distributed computing, other fields related to wireless networks, information theory, and coding theory.
\end{biography}
\begin{biography}[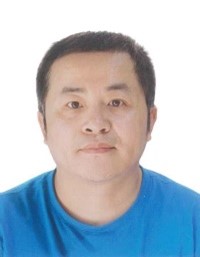]
\noindent
\textbf{Bin Dai} received the B.Sc. degree in communications and information systems from the University of Electronic Science and Technology of China, Chengdu, China, in 2004, and the M.Sc. and Ph.D. degrees in computer science and technology from Shanghai Jiao Tong University, Shanghai, China, in 2007 and 2012, respectively. In 2011 and 2012, he was a Visiting Scholar with the Institute for Experimental Mathematics, Duisburg-Essen University, Essen, Germany. In 2018 and 2019, he was a Visiting Scholar with the Department of Electrical and Computer Engineering, Ohio State University (OSU), Columbus, USA. He is currently a Professor with the Southwest Jiaotong University. His research interests include information-theoretic security and network information theory and coding.
\end{biography}
\begin{biography}[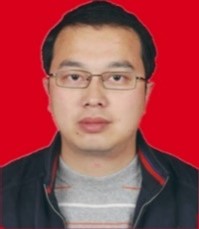]
\noindent
\textbf{Xiaohu Tang} received the B.S. degree in applied mathematics from Northwest Polytechnic University, Xi\text{'}an, China, in 1992, the M.S. degree in applied mathematics from Sichuan University, Chengdu, China, in 1995, and the Ph.D. degree in electronic engineering from Southwest Jiaotong University, Chengdu, in 2001. From 2003 to 2004, he was a Research Associate with the Department of Electrical and Electronic Engineering, Hong Kong University of Science and Technology. From 2007 to 2008, he was a Visiting Professor with the University of Ulm, Germany. Since 2001, he has been with the School of Information Science and Technology, Southwest Jiaotong University, where he is currently a professor. His research interests include coding theory, network security, distributed storage, and information processing for big data.

Dr. Tang was a recipient of the National Excellent Doctoral Dissertation Award in 2003, China, the Humboldt Research Fellowship in 2007, Germany, and the Outstanding Young Scientist Award by NSFC in 2013, China. He served as Associate Editors for several journals, including IEEE Transactions On Information Theory and IEICE Transactions on Fundamentals, and served on a number of technical program committees of conferences.
\end{biography}

\mbox{}
\clearpage
\clearpage
\large
\noindent

\end{document}